\documentclass{jfm}

\usepackage{graphicx}
\usepackage{subcaption}
\usepackage{caption}
\usepackage{newtxtext}
\usepackage{newtxmath}
\usepackage{natbib}
\usepackage{hyperref}
\hypersetup{
    colorlinks = true,
    urlcolor   = blue,
    citecolor  = black,
}

\newcommand{\RomanNumeralCaps}[1]
\linenumbers

\usepackage{epsfig}
\usepackage{color}

\usepackage{psfrag}

\newcommand\be{\begin{equation}}
\newcommand\ee{\end{equation}}
\newcommand\q{\quad}

\newcommand{\dt}{\dot{d}}
\newcommand{\ddt}{\ddot{d}}
\newcommand\C{\mathcal{C}}
\newcommand\F{\mathcal{F}}
\newcommand\G{\mathcal{G}}

\newcommand{\tA}{{\bf{\mathsf{A}}}}

\newcommand\oC{\overline{C}}
\newcommand\hC{\hat{C}}
\newcommand\hk{\hat{k}}
\newcommand\hb{\hat{b}}
\newcommand\hS{\hat{S}}
\newcommand\hA{\hat{A}}

\title{Aquatic locomotion by an elastically mounted  flexible foil actuated by an oscillating force}

\author{R. Fernandez-Feria\aff{1}\corresp{\email{ramon.fernandez@uma.es}}
}
\affiliation{\aff{1}{Fluid Mechanics Group, IMEC.UMA, University of M\'alaga, Dr Ortiz Ramos s/n, 29071 M\'alaga, Spain}}

\begin{document}

\maketitle

\begin{abstract}

An analytical formulation of the fluid-structure interaction of a flexible foil driven by an oscillating force actuating on its elastically  mounted leading edge, so that it can heave, pitch and deform passively with the hydrodynamic forces,  is used to investigate  the  aquatic locomotion of a body, responsible for the whole drag and thrusted by the  oscillating flexible foil. The small-amplitude theoretical model is validated with previous theoretical and experimental results for a body propelled by a rigid plate oscillating with a prescribed heaving motion and  passive pitch. The inclusion of passive heave and deformation  allows to expand the parametric ranges for optimal self-propulsion conditions in terms of length travelled by flapping cycle (stride length) and locomotion efficiency. In addition to the known optimal locomotion condition localized near the resonance of the torsional spring on which the foil is elastically mounted, which  here is modulated by its coupling with the resonances of the translational spring and of the structural deformation of the foil, another even better local optimal locomotion condition is found near the translational spring branch  of the elastic support resonance that occurs at lower stiffnesses of both springs. Unlike the local maximum of efficiency close to the natural frequency associated with the torsional spring branch, which increases with the stiffness of the foil, being the highest for a rigid foil, the larger local maximum associated with  the translational spring branch increases as the stiffness of the foil decreases.

\end{abstract}

\begin{keywords}
flow-structure interactions, propulsion, swimming/flying\end{keywords}

\section{Introduction}
\label{sec_intro}

Inspired by the efficient propulsion of many aquatic animals that produce most of their thrust by the oscillatory motion of their caudal fins, the study of flapping foils to propel aquatic vehicles has grow enormously in recent years \citep{laude15,zhuwh19,prana24,yaoxi25}. Among the different strategies analyzed  to enhance thrust and propulsive efficiency generated by pitching and heaving foils,  passive pitch allowed by  localized leading edge flexibility 
in a rigid foil actively heaved in a fluid  has been shown to generate a large  thrust peak  when the oscillating foil is actuated at frequencies close the resonance associated with the stiffness of its torsional spring  support \citep{spamo10,moore14}. Moreover, several studies have also showed that the elastic  properties of the peduncle joining  the caudal fin to the trunk, and their tuning, are closely tied to the optimal  thrust production and  propulsive efficiency in the locomotion of many aquatic animals \citep{laude00,fisla06,lucjo14,zhozh21}.

We focus here on the aquatic locomotion of animals or bioinspired robots which can be simply modelled as a virtual rigid body, carrying most of the mass and drag, self-propelled by a rear oscillating foil which generates the necessary thrust  \citep[e.g.][]{akomo18,mooqu19,sanra20,panpa21,grapa25,grapa26}, specially when this model, which for short is designated VB-OF model (for virtual body-oscillating foil), allows for an analytical solution of the  flow-structure interaction.  The kind of self-propulsion modelled by the VB-OF model is actually quite different from the undulatory locomotion of most aquatic animals that generate a travelling wave along their flexible body or propulsive fins to push fluid backwards and thus produce the necessary thrust \citep{smits19}, or use a mixed type of locomotion between the purely undulatory mode and that idealized by the VB-OF model. But aquatic animals whose locomotion mode approaches that of the VB-OF model, such as   tunas and bonitos, are among the fastest and most efficient swimmers in sustained cruise \citep{webb84}, and they are endowed by nature with a long, almost rigid-fin tail, thus approximating better the further assumption of a two-dimensional foil  used  in most theoretical works on oscillatory aquatic locomotion since the pioneering work by  \cite{light70}.

Particularly, we focus on models where the oscillating foil is elastically connected to the virtual body, simulating the flexibility of the muscles that move laterally or vertically the tail of some aquatic animals, so that  a complete or partially passive oscillation of the fin is allowed to generate, as mentioned above, a very noticeable increase in thrust and propulsive efficiency. Although many of the previous works analyzing this mechanism to enhance the propulsion efficiency taking advantage of the resonances of the elastic joint assume that the foil is immersed in a uniform current with a given constant speed \citep[e.g.][]{wilis07,zhali10,bocst14,moore14,moore15,thaph18,photh20,feral21a,meiya23,huyan25}, here we focus on self-propelled locomotion, so that the swimming speed is not given but  an additional unknown of the problem, obtained from the balance between the drag of the virtual body and  the thrust generated by the oscillating foil.  The swimming velocity is  therefore obtained along the other unknowns of the problem related to the passive motion and deformation of the foil.

In this context, \cite{zhozh21} developed a VB-OF self-propulsion model to explain their   experimental results using a tuna-like platform that tunes its own tail stiffness using a motor-driven `muscle'. Their experiments showed that to maximize efficiency, muscle tension should scale with swimming speed squared, offering a simple tuning strategy for this kind of fish-like robots. The  VB-OF model developed by these authors, based on linear potential-flow theory treating the tail fin as a two-dimensional rigid foil whose heaving motion is prescribed (that generated by the experimental actuator) and whose pitch angle responded passively on the basis of hydrodynamic forces and the torsional spring tension, remarkably explained and reproduced their experimental results. It will be shown below that this model developed by \cite{zhozh21} is a particular case of the more general one developed here, and therefore their results coincide with the ones reported here in a limiting case.  

More recently, \cite{grapa26} presented a similar  VB-OF analytical  self-propulsion model  to that developed  by \cite{zhozh21} for a rigid foil elastically mounted at its leading edge through a torsional spring, discussing the best performance in terms of the stride length and efficiency as a function of an appropriate non-dimensional frequency, and   analyzing the role of fluid damping on the optimal locomotion conditions.The results by these authors are also reproduced and discussed below as a limiting  case of the  analytical formulation developed in the present study.

Here we consider an analytical  VB-OF self-propulsion model with a flexible foil elastically mounted on its leading edge through  torsional and  translational springs and dampers, thus allowing for both passive heave and pitch, in additional to the passive flexural deformation of the foil, and actuated by a driving oscillating force that simulates the muscle force that generates the lateral (or vertical)  tail motion, instead of assuming a  prescribed heaving motion in a rigid foil  like in the previously mentioned studies. The inclusion of passive heave and deformation  allows to expand the parametric ranges  to search analytically  for the optimal locomotion conditions. The model  presented here complements that reported  in \cite{lopfe23a}, which considered a similar VB-OF self-propulsion model but with the foil driven by an oscillating torque instead of an actuating force that better reproduces the lateral force exerted by the peduncle that joins the fish's tail to its trunk. Moreover, here we use a foil deformation model with additional degrees of freedom, which extends the validity of the analytical solutions of the flow-structure interaction to  much lower values of the  stiffness of the plate \citep{ferna25}. Though both models share partially some results, especially some related to the springs resonances, the present one yields  new, more straightforward analytical results because it consider directly the steady state locomotion instead of the whole temporal evolution of the self-propelled body. Thus, the results for the parametric ranges for optimal locomotion performance presented here in terms of the stride length and propulsion efficiency are new. These results are, in addition,  discussed  in relation to, and validated with, the results of the previous works for a rigid foil with prescribed heave cited above, which include experimental self-propulsion results.

\begin{figure}
  \begin{psfrags}
  \psfrag{(a)}[c][l][1.2]{(a)}
  \psfrag{(b)}[c][l][1.2]{(b)}
   \psfrag{U}[l][l][1]{$U$}
    \psfrag{x}[l][l][1]{$x$}
      \psfrag{h}[l][l][1]{$h(t)$}
        \psfrag{a}[l][l][1]{$\alpha(t)$}
    \psfrag{e}[l][l][1]{$\varepsilon$}
    \psfrag{c}[l][l][1]{$c$}
    \psfrag{kh}[l][l][1]{$k_h, b_h$}
    \psfrag{ka}[l][l][1]{$k_a, b_a$}
    \psfrag{z}[l][l][1]{$z_s(x,t)$}
 \psfrag{Li}[c][l][1]{$L_{i0} \cos(\omega \tilde{t} )$}
 \psfrag{Vi}[c][l][1]{Virtual body}
  \psfrag{CD}[l][l][1]{$\overline{C}_D$}
  \psfrag{CT}[l][l][1]{$\overline{C}_T$}
   \centerline{\epsfig{file=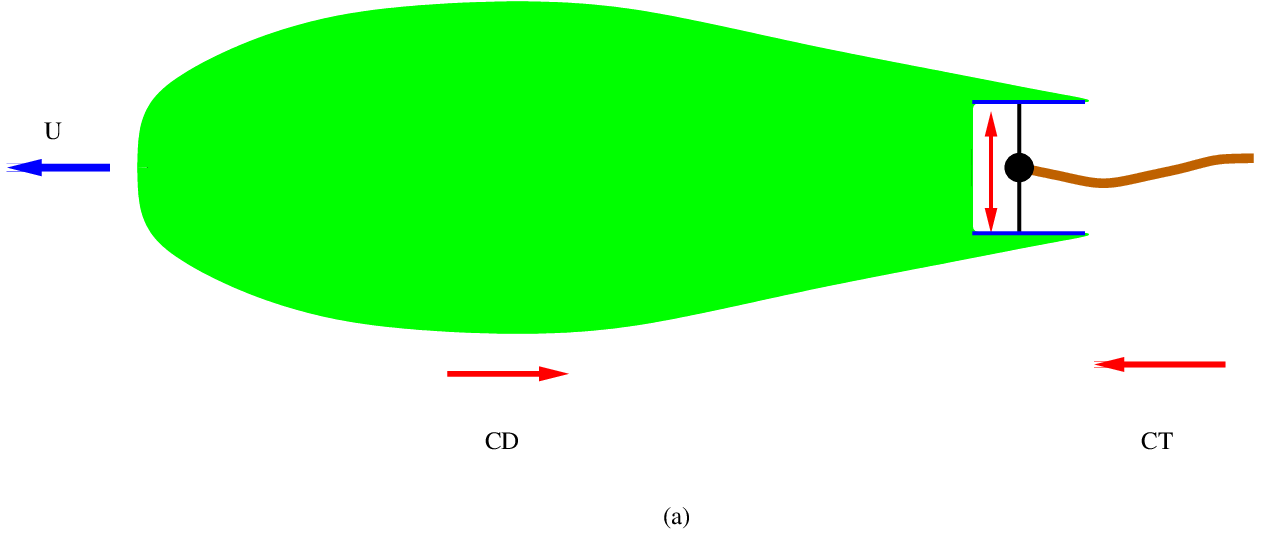,width=0.55\linewidth}\hspace{10mm}\epsfig{file=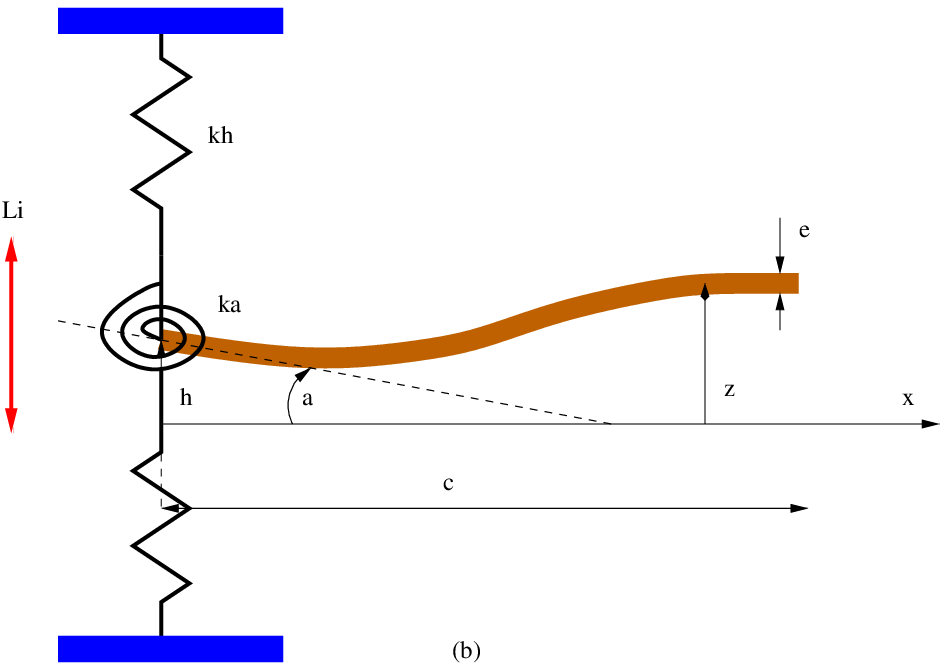,width=0.45\linewidth}}
  \end{psfrags}
  \caption{Schematic of the problem with the VB-OF self-propulsion model.}
  \label{fig_sketch}
\end{figure}

\section{Formulation of the problem}
\label{sec_for}

We study the locomotion of an aquatic animal or robot that can be modeled by the combination of a virtual body responsible for the entire drag and a tail, consisting of a two-dimensional flapping foil, which produces  the thrust. Self-propulsion, or cruising, velocity $U$ is achieved through the balance between body drag and tail thrust (see Fig. \ref{fig_sketch}a). As described in the Introduction, this simple VB-OF model  has been widely used to understand and to optimize aquatic bio-inspired  locomotion. Here, the foil is considered flexible and elastically mounted to the virtual body through springs and dampers at its leading edge, so that the heaving and pitching motions are both passive when actuated by a given oscillatory transversal  force (see Fig. \ref{fig_sketch}b). The deformation of the foil is also obtained by solving analytically the fluid-structure interaction (FSI). Once the passive oscillation and deformation of the foil are obtained, the thrust force can also be computed analytically, and hence the cruising velocity. Since the actuating force  is given, the input power, and therefore the propulsive efficiency, are also straightforwardly  computed (see \S \ref{sec_perform}).

The analytical expressions  for the fluid  force and moments derived in \cite{ferna25}, necessary to solve the FSI, are used here assuming that the foil transversal displacement $z=z_s(x,t)$ is small compared to its chord length $c$ and that its instantaneous shape is approximated   by a fifth-order polynomial in the stream-wise coordinate $x$. In non-dimensional form,  with $x$ and $z$ scaled with  $c/2$ and time $t$  scaled with $c/(2U)$ (remember that the cruising velocity $U$ is here unknown),  it can be written as:
\[ z_s(x,t)= h(t) - \alpha(t) (x+1) + d_1(t) [ 24(x+1)^2-8(x+1)^3+(x+1)^4] \]
\be + d_2(t)[160(x+1)^2 -40(x+1)^3+(x+1)^5] \,, \q\q -1 \leq x \leq 1 \,, 
\label{zs_def}
\ee 
where $h(t)$ and $\alpha(t)$ characterize the (unknowns) heaving and pitching motions of the pivot axis located at  the leading edge ($x=-1$), and $d_1(t)$ and $d_2(t)$ are two additional unknown amplitudes characterizing the passive flexural deformation of a foil with a free trailing edge, $(\partial^2 z_s/\partial x^2)_{x=1}=$$(\partial^3 z_s/\partial x^3)_{x=1}=0$. This approximation covers up to the second bending natural  frequency of the fluid-foil system, and it was shown  to provide accurately the FSI  for  values of the non-dimensional stiffness $S$ (defined below)  from infinity, corresponding to a rigid foil,  down to order of $10^{-1}$.

The balances  of the  lateral (or vertical)  forces on the foil   and of  their moments about its leading edge can be obtained by integrating  the two-dimensional Euler-Bernoulli beam equation along its chord length multiplied by $1$ and $1+x$, respectively, when the  displacement \eqref{zs_def} is used, to obtain \citep[][everything  is non-dimensional]{ferna25}:
\be R \left( \ddot{h} - \ddot{\alpha} + \frac{96}{5} \ddot{d}_1 + \frac{416}{3} \ddot{d}_2 \right) =C_L+C_{L_i}-k_h-b_h \dot{h} \,, \label{mom1a}
\ee
\be R \left( \frac{1}{2} \ddot{h} - \frac{2}{3} \ddot{\alpha} + \frac{208}{15} \ddot{d}_1 + \frac{704}{7} \ddot{d}_2 \right) =C_M+C_{M_i}+k_a \alpha + b_a \dot{\alpha} \,, \label{mom2a}
\ee
where dots are used for the derivatives with respect to the non-dimensional time $t$, $C_L$ and $C_M$ 
are the lift and moment coefficients, i.e. the lift force and moment exerted by the fluid per unit span of the foil  scaled with $\rho U^2 c/2$ and $\rho U^2 c^2/2$, respectively, with $\rho$ the fluid density; $C_{L_i}$ and $C_{M_i}$ are the input point force and torque at the leading edge that generate the motion of the foil, both non-dimensionalized in the same way as $C_L$ and $C_M$; 
$k_h$,  $b_h$, $k_a$ and $b_a$  are the non-dimensional constants characterizing the translational spring and damper, and the torsional spring and damper, respectively, through which the foil is elastically mounted on its leading edge, related to their dimensional counterparts  through  (dimensional quantities are written   here with a tilde,  ` $\tilde{\hspace{2mm}}$', when the same symbol is used for the corresponding dimensionless quantity)
\be k_h = \frac{\tilde{k}_h}{\rho U^2} \,, \q b_h = \frac{\tilde{b}_h}{\rho U c /2} \,, \q k_a = \frac{\tilde{k}_a}{\rho U^2 c^2/2} \,, \q b_a = \frac{\tilde{b}_a}{\rho U c^3 /4} \,; \label{springs}
\ee
finally,
\be R = \frac{4 m}{\rho c^2} = \frac{4 \rho_s \varepsilon}{\rho c} \,, \label{def_R}
\ee
is the mass ratio, where $m = \rho_s \varepsilon c$ is the mass of the foil per unit span, with $\varepsilon$  the foil thickness and $\rho_s$  the solid density. 

Note in Eq. \eqref{mom2a} that $2 R/3$ is the non-dimensional moment of inertia of the plate about its leading edge, and that the moments are taken positive when counterclockwise,  whereas the pitch angle $\alpha$ is assumed positive when clockwise, as it is usual in aerodynamics. For a rigid foil ($d_1=d_2=0$), Eqs. \eqref{mom1a}-\eqref{mom2a} are the well known  equations of motion of a plate elastically mounted in the small-amplitude limit. 

For a flexible foil, there are four unknown quantities, $h(t)$, $\alpha(t)$, $d_1(t)$ and $d_2(t)$, and one needs two additional equations to obtain the deformation of the foil through $d_1$ and $d_2$. They are derived from the following two moments of the Euler-Bernoulli beam equation; i.e., by multiplying it by $(x+1)^2$ and $(x+1)^3$, respectively, and integrating between $x=-1$ and $x=1$ \citep{ferna25}:
\be R \left( \frac{4}{3} \ddot{h} - 2 \ddot{\alpha} + \frac{4544}{105} \ddot{d}_1 + \frac{944}{3} \ddot{d}_2 \right) + S \left( \frac{32}{3} d_1 + 80 d_2 \right) = C_{F_1}  \,, \label{mom3a}
\ee
\be R \left( 2 \ddot{h} - \frac{16}{5} \ddot{\alpha} + \frac{496}{7} \ddot{d}_1 + \frac{32 512}{63} \ddot{d}_2 \right) + S \left( 16 d_1 + 128 d_2 \right) = C_{F_2}  \,, \label{mom4a}
\ee
where $C_{F_1}$ and $C_{F_2}$ are the first two flexural moments about the leading edge exerted by the fluid on the foil. 
The non-dimensional bending stiffness parameter has been defined as
\be S= \frac{48 EI }{\rho U^2 c^3} = \frac{4 E \varepsilon^3}{\rho U^2 c^3} \,, \label{def_S}
\ee
where the structural bending rigidity of the foil per unit span, $E I = E \varepsilon^3/12$, has  been assumed constant.

In the present work we consider the locomotion generated by the action of a given oscillatory force at the foil's leading edge with frequency $\omega= 2 \pi f$; i.e., when $C_{M_i}=0$ and $C_{L_l} \equiv  L_i /(\rho U^2 c/2)$ is given by the real part of
\be C_{L_i} \equiv \frac{L_{i0} e^{i \omega \tilde{t}}}{\rho U^2 c/2}=  A_l e^{i \omega \tilde{t}} = A_l e^{ik t}  \,, \label{inlift}
\ee
where $A_l$ is the non-dimensional intensity (or amplitude) of the applied oscillatory force. The non-dimensional o reduced frequency is
\be k = \frac{\omega c}{2 U} = \frac{ \pi f}{U} \,. \label{red_fre}
\ee
Actually, $U$   is the main unknown sought in the self-propulsion problem, whose more convenient dimensionless form is related to the inverse of the reduced frequency $k$,
\be S_L = \frac{U}{c f} = \frac{\pi}{k} \,, \label{def_SL}
\ee
which is the stride length, or chord lengths travelled per flapping cycle. 

With this harmonic forcing it is expected that the foil eventually acquires a harmonic displacement given by \eqref{zs_def} with
\be h(t)= h_0 e^{i k t} \,, \q \alpha(t)= \alpha_0 e^{ikt} \, \q  d_1(t)= d_{10} e^{i k t} \,, \q d_2(t)= d_{20}  e^{ikt} \,, \label{harmo}
\ee
where $h_0$, $\alpha_0$, $d_{10}$ and $d_{20}$ are unknown complex quantities. For this  harmonic motion and deformation, the fluid force and moments coefficients $C_L$, $C_M$, $C_{F_1}$ and $C_{F_2}$ can be obtained analytically \citep{ferna25}, and are given in Appendix \ref{app_coeffi}.  For a rigid foil ($d_1=d_2=0$) the expressions for the lift and moment coefficients, $C_L$ and $C_M$,  are the well-known classical ones derived by \cite{theod35}. 

Substituting these relations into Eqs. \eqref{mom1a}, \eqref{mom2a}, \eqref{mom3a} and \eqref{mom4a}, and for a given value of $k$, we get a system of four lineal equations  for $h_0$, $\alpha_0$, $d_{10}$ and $d_{20}$ that can be written in the form
\be  \tA  \cdot {\bf{d}}_0 \equiv (\tA_0+\tA_f)  \cdot {\bf{d}}_0 = {\bf{b}} \,, \q\q \q   {\bf{d}}_0  = \left( \begin{array}{c} h_0\\ \alpha_0 \\ d_{10}\\ d_{20}
                                                          \end{array} \right) \,,  
\q\q \q   {\bf{b}}  = \left( \begin{array}{c} A_l\\ 0 \\ 0\\ 0
                                                          \end{array} \right) \,, \label{def_eq_Ad0}
\ee
where the matrix $\tA_0$ within $\tA=\tA_0+\tA_f$  takes into account the passive motion and deformation of the foil \em  in vacuum\rm, i.e. the inertia of the foil and its elastic support, whereas the matrix $\tA_f$ accounts for the fluid-structure interaction through $C_L$, $C_M$, $C_{F_1}$ and $C_{F_2}$. The (complex)  terms of both matrices are given in Appendix \ref{app_coeffA}. The elements of $\tA_0$ depend on the non-dimensional parameters $k$, $R$, $S$, $k_h$,  $b_h$, $k_a$ and $b_a$, while $\tA_f$ only depends on $k$. Obviously,  the solution ${\bf{d}}_0 $  is independent of $A_l$ when scaled with $A_l$ due to the linearity of the system.

Following previous works using a similar self-propulsion model, but for a rigid foil \citep[e.g.][]{mooqu19,panpa21,grapa26},   to obtain $k$, and therefore the cruising velocity $U$ for a given frequency $f$ and a given set of non-dimensional parameters, one uses the balance between the time-averaged thrust and drag,
\be \oC_D = \oC_T \,. \label{CDCT}
\ee
The time-averaged thrust for the foil displacement \eqref{zs_def} with the complex harmonic amplitudes \eqref{harmo} was also obtained, and checked with available experimental data,  in \cite{ferna25} (the expression is given in  Appendix E of that reference). For the body drag $ \oC_D$ we shall assume a given constant value, as it is done in the previous works cited.

The solution  ${\bf{d}}_0$ of Eq. \eqref{def_eq_Ad0} can be written, in amplitude and phase, as  
\be h_0=h_{m} e^{- i \varphi} \,, \q \alpha_0 = a_0 e^{i \phi} \,, \q d_{10}=d_{1m}e^{i\psi_1} \,, \q d_{20} = d_{2m} e^{i \psi_2} \,. \label{harmo1}
\ee
Since the expression derived for $\oC_T$  takes as a reference for the phase shift that of the heaving motion, i.e., with $\varphi=0$ in Eq. \eqref{harmo1}, the phase shifts in \eqref{harmo1} are redefined by adding an angle $\varphi$ to all of them, which due to the linearity of \eqref{def_eq_Ad0} is physically equivalent to a foil motion generated by an input force 
\be C_{L_i} = A_l e^{i(kt+\varphi)} \,.  \label{ampli2} \ee
This has to be taken into account when computing the input power (see \S \ref{sec_perform} below). 

\subsection{Solution procedure. Redefinition of some the parameters}
\label{sec_sol_pro}
By jointly solving the set of  linear equations \eqref{def_eq_Ad0} and the equation \eqref{CDCT}, the reduced frequency $k$ and the four complex quantities \eqref{harmo1} are obtained. Since this is a non-linear problem in $k$, some iterative process in the parameter $k$ is needed to reach a solution (see below). As mentioned above, $U$ is then obtained   from  $k$ using its definition \eqref{red_fre}. But, for the iterative process to be possible, it is convenient that all the other non-dimensional parameters in Eq. \eqref{def_eq_Ad0} do not contain $U$. For this reason, the remaining dimensionless parameters, except $R$,  are redefined in the following way (marking them with a hat `$\hat{\hspace{2mm}}$'):
\be  \hS = \frac{16 E \varepsilon^3}{\rho \omega^2 c^5} = \frac{S}{k^2}  \,, \q \hk_h = \frac{4 \tilde{k}_h}{\rho \omega^2 c^2} = \frac{k_h}{k^2} \,, \q \hb_h = \frac{4 \tilde{b}_h}{\rho \omega c^2} = \frac{b_h}{k}  \,,  \label{paramo1}
\ee
\be \hk_a = \frac{8 \tilde{k}_a}{\rho \omega^2 c^4} = \frac{k_a}{k^2}  \,, \q \hb_a = \frac{8\tilde{b}_a}{\rho \omega c^4} = \frac{b_a}{k} \,. \q    \label{paramo2}
\ee
When these expressions are substituted into Eqs. \eqref{mom1a}, \eqref{mom2a}, \eqref{mom3a} and \eqref{mom4a} for the harmonic motion \eqref{harmo}, after dividing by $k^2$ and writing the input lift \eqref{inlift} as
\be C_{L_i} \equiv \hC_{L_i} k^2 = \hA_l k^2 e^{i k t } \,, \q \text{with} \q \hA_l = \frac{8 L_{i0}}{\rho \omega^2 c^3} \,, \label{inlift2}
\ee
the system  of linear equations \eqref{def_eq_Ad0} can be written as
\be   \hat{\tA}  \cdot {\bf{d}}_0 \equiv (\hat{\tA}_0+\hat{\tA}_f)  \cdot {\bf{d}}_0 = {\hat{\bf{b}}} \,, \label{sistema}
\ee
with ${\hat{\bf{b}}}= (\hA_l,0,0,0)^T$, and the re-scaled matrices
\be  \hat{\tA} = \frac{\tA}{k^2} \,, \q  \hat{\tA}_0 =\frac{\tA_0}{k^2} \,, \q  \hat{\tA}_f=\frac{\tA_f}{k^2} \,, \label{matri}
\ee
with $S$, $k_h$,  $b_h$, $k_a$ and $b_a$ replaced by their hatted counterparts. In this way, $U$ only enters in the system \eqref{sistema} (and, of course, in $\oC_T$)  through the non-dimensional parameter $k$. Actually, the new, hatted inertial matrix $\hat{\tA}_0$ is independent of $k$ (see Eq. \eqref{def_A0hat} in Appendix \ref{app_coeffA}.).

The problem thus formulated with Eqs. \eqref{sistema} and \eqref{CDCT} depends on the non-dimensional parameters $R$, $\hS$, $\hk_h$,  $\hb_h$, $\hk_a$, $\hb_a$ and $\oC_D$ to yield the unknowns \eqref{harmo1}, and $k$, which provides  the stride length  $S_L$ through \eqref{def_SL}. However, the original system \eqref{def_eq_Ad0}, and more precisely the matrix $\tA$, is also needed to obtain the non-dimensional natural frequency $k_r$ by minimizing $|\det(\tA)|$, because in that notation the frequency $\omega$ only enters in the non-dimensional parameter $k$. This non-dimensional natural frequency is useful for, among other things, representing the results in terms of the ratio between the actuating frequency $\omega$ and the corresponding natural frequency $\omega_r$ for the given set of non-dimensional parameters,
\be \frac{\omega}{\omega_r} = \frac{k}{k_r} \,. \label{sigr}
\ee
In the present self-propulsion problem, this ratio is also part of the solution, but sometimes it is convenient to know how the other unknowns are related to the resonant frequency of the system rather than in terms of the input parameters.

The computing procedure for a given set of the governing non-dimensional parameters, that is, for given $R$, $\hS$, $\hk_h$,  $\hb_h$, $\hk_a$, $\hb_a$ and $\oC_D$, can be summarized as follows:  (i) Start assuming  a (low) value of $k$ and solve \eqref{sistema} to obtain \eqref{harmo1}; (ii) with these values compute $\oC_T$; (iii) if $\oC_T$ is smaller than the selected value of $\oC_D$, increase $k$ by a small quantity; repeat the procedure until the balance \eqref{CDCT} is achieved. This will provide the stride length \eqref{def_SL}, and all the other self-propulsion performance parameters described in the next section, for the given set of non-dimensional governing parameters.

\section{Performance parameters}
\label{sec_perform}

In addition to the stride length, or non-dimensional cruising speed \eqref{def_SL}, which is arguably the most relevant searched parameter, another important performance parameter is the locomotion efficiency,  related to the input power.

Once the passive motion of the foil is obtained by solving Eq. \eqref{sistema}, particularly, the passive heave amplitude  $h_0$, the non-dimensional input power (per unit span) needed to generate it can be computed from
\be C_P(t) \equiv \frac{P_i}{\rho U^3 c/2}  = \Re[\dot{h} ] \Re[ C_{L_i}] = \Re[ik h_0 e^{ikt}] \Re[\hA_l k^2 e^{i(kt+\varphi)}]  
\,. \label{cp1} \ee
Its time-averaged  over a cycle is 
\be \oC_P = \frac{1}{2} k^3 h_0 \hA_l \sin \varphi  
\,. \label{cp2} \ee
One of the main advantages of this configuration with respect to previous studies based on prescribed heave or pitch, instead of the present prescribed input force and passive motion, is the simplicity of this expression for the input power, which is also physically  more interesting because it depends only on the applied point force and the passive oscillation that it generates. It incorporates implicitly  the power that the fluid exerts on the oscillating flexible foil, including the effects of inertia and deformation, through the solution of \eqref{sistema}, which yields the passive heave amplitude $h_0$.

As a measure of the locomotion efficiency one may use the ratio of the obtained cruising speed and the input power, $U/P_i$, or its inverse, related to the so-called cost of transport \citep[e.g.][]{panpa21,zhozh21}. However, these are dimensional quantities. For that reason, it is preferred here the use of a dimensionless form of the first of these quantities, the ratio of the stride length and the input power coefficient,
\be 
E_f = \frac{S_L}{\oC_P} \,. \label{EFF}
\ee
Though dimensionless, this quantity is not an efficiency in the strict sense because it can be larger than unity. In any case, as we shall see in the reported results, $S_L$ and $E_f$  are always closely related, 
so that $S_L$ is usually enough to characterize the swimming efficiency.

Another relevant non-dimensional parameter that characterizes flapping foil self-propulsion is the Strouhal number, defined as the frequency $f$ multiplied by the beat amplitude (maximum peak-to peak trailing edge amplitude, $|A|= |z_s(x=1)| c$,  divided by the cruising velocity $U$:
\be \text{St} = \frac{f |A|}{U} = \frac{|h_0-2 \alpha_0+48 d_{10}+ 352 d_{20}|}{S_L} \,. \label{strouhal}
\ee
Experimental data on the flapping propulsion of many  aquatic animals show that this parameter remains in a narrow range, between approximately $0.2$ and $0.4$,  in the optimal cruise swimming \citep{tritr93,taynu03,eloy12}.

\section{Locomotion by a rigid foil elastically mounted}
\label{sec_rigid}

\subsection{Comparison  with previous works with prescribed heave. Validation of the model}
\label{sec_pre_rig}
For a rigid foil ($S \to \infty$), Eqs. \eqref{mom3a}-\eqref{mom4a} yield $d_{1}=d_{2}=0$, and one has to solve only Eqs. \eqref{mom1a}-\eqref{mom2a} for the passive heave and pitch. This self-propulsion problem has been considered very recently by \cite{grapa26}, but assuming that the heave is prescribed, instead of been also generated passively by an input force, so that one has only to solve equation \eqref{mom2a} for the pitch $\alpha(t)$ for a given $h(t)$:
\be R \left( \frac{1}{2} \ddot{h} - \frac{2}{3} \ddot{\alpha} \right) =C_M+k_a \alpha + b_a \dot{\alpha} \,. \label{mom2b}
\ee
On using Theodorsen's moment for $C_M$ (i.e., Eq. \eqref{CM0} with \eqref{G0t} and $d_1=d_2=0$) and solving for $\alpha(t)$ (grouping the terms with $\alpha$ on the right-hand side of the equation),
\be \left( \frac{2 R}{3}  + \frac{9 \pi}{16}  \right) \ddot{\alpha} - \left[ \frac{3\pi}{4} \left(1 + \C(k) \right) + b_a \right] \dot{\alpha} - \left(\frac{\pi \C(k)}{2} + k_a \right) \alpha =  \left( \frac{R}{2}  + \frac{\pi}{2}  \right) \ddot{h} + \frac{\pi \C(k)}{2} \dot{h} \,, \label{alpharig1}
\ee
where $\C(k)=\F(k)+i\G(k)$ is Theodorsen's (complex) function \citep[e.g.][]{garri36}. Inserting the harmonic motion \eqref{harmo} and using the non-dimensional ({\it{hatted}}) parameters \eqref{paramo2}  that are independent of $U$ -- i.e., using the second row of the system \eqref{sistema} -- one can obtain explicitly the pitch amplitude and phase in relation to the given heave amplitude $h_0$:
\be \alpha_0 = a_0 e^{i\phi} = \frac{\left[\frac{R}{2} + \pi \left(\frac{1}{2} - \frac{i  \C(k)}{2k} \right) \right] h_0}
	{\frac{2 R}{3} - \hk_a - i \hb_a - \pi \left[ \frac{3 i}{4 k} - \frac{9}{16} + \C(k) \left( \frac{3i}{4 k} + \frac{1}{2 k^2} \right) \right] }  \,. \label{alpharig2}
\ee
Note that the terms multiplied by $\pi$ come from the FSI, the terms with $R$ from the inertia of the foil and the remaining terms from the torsional spring and damper. It is checked that $\alpha_0$ from \eqref{alpharig2} obviously coincide with the solution of the whole system \eqref{sistema} when very large values of $S$ and $\hk_h$ are used, together with the input lift amplitude $\hA_l=\hk_h h_0$, so that the first equation yields directly a prescribed heave amplitude $h_0$, without phase shift, the third and fourth $d_{10}=d_{20}=0$, and the second one reproduces exactly \eqref{alpharig2}.

This expression \eqref{alpharig2} coincides with Eq. (18) in \cite{grapa26}, except that these authors include no damper ($\hb_a=0$) and their  pitch has the opposite sign (here  $\alpha$ is positive when clockwise, as described in \S \ref{sec_for}). This expression is also contained in the analyses of \cite{moore14} and \cite{feral21a} for a given free-stream velocity $U$ (i.e., for given $k$). From the real part of the denominator of \eqref{alpharig2}, the resonant frequency in terms of the non-dimensional parameter $\hk_a$  defined in \eqref{paramo2} is
\be \hk_{ar} = \frac{2 R}{3} + \frac{9 \pi}{16} +  \frac{3 \pi \G(k)}{4 k} - \frac{\pi \F(k)}{2 k^2} \,,
\ee
or dimensionally, on using \eqref{paramo2},
\be \omega_r = \sqrt{\frac{8 \tilde{k}_a}{\rho c^4 \left(  \frac{2 R}{3} + \frac{9 \pi}{16} +  \frac{3 \pi \G(k)}{4 k} - \frac{\pi \F(k)}{2 k^2} \right)}}  \,,
\ee
which obviously coincides with that reported by \cite{grapa26}. In the limit $k \gg 1$ ($S_L \ll 1$), $\omega_r \simeq \sqrt{ 8 \tilde{k}_a/\left[\rho c^4 \left(  \frac{2 R}{3} + \frac{9 \pi}{16} \right)\right]}$, an expression already reported by \cite{moore14}, among others, which contains the inertia contribution (term with $R$) and the leading FSI term.

\begin{figure}
\centerline{\epsfig{file=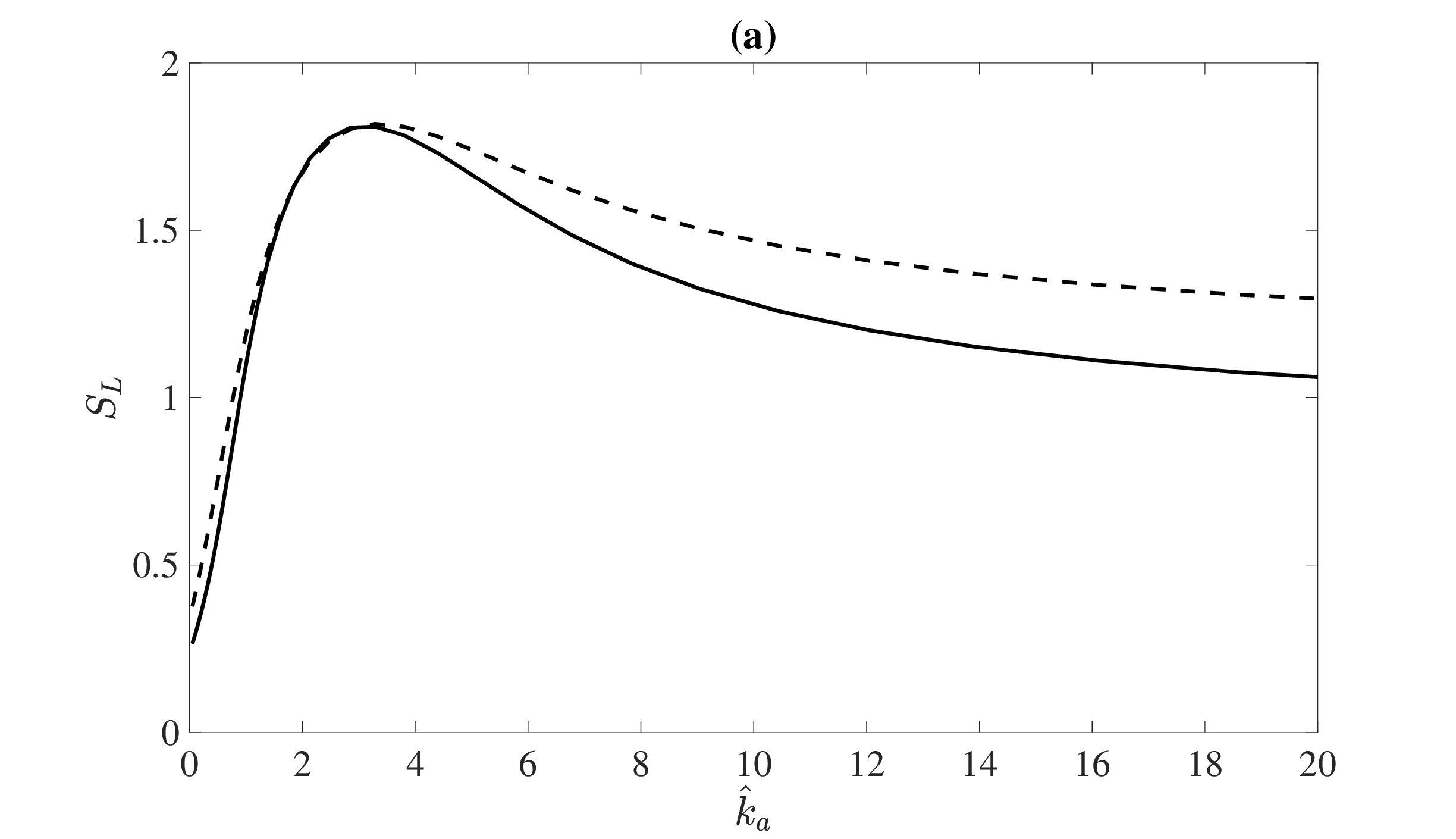,width=.55\linewidth}\epsfig{file=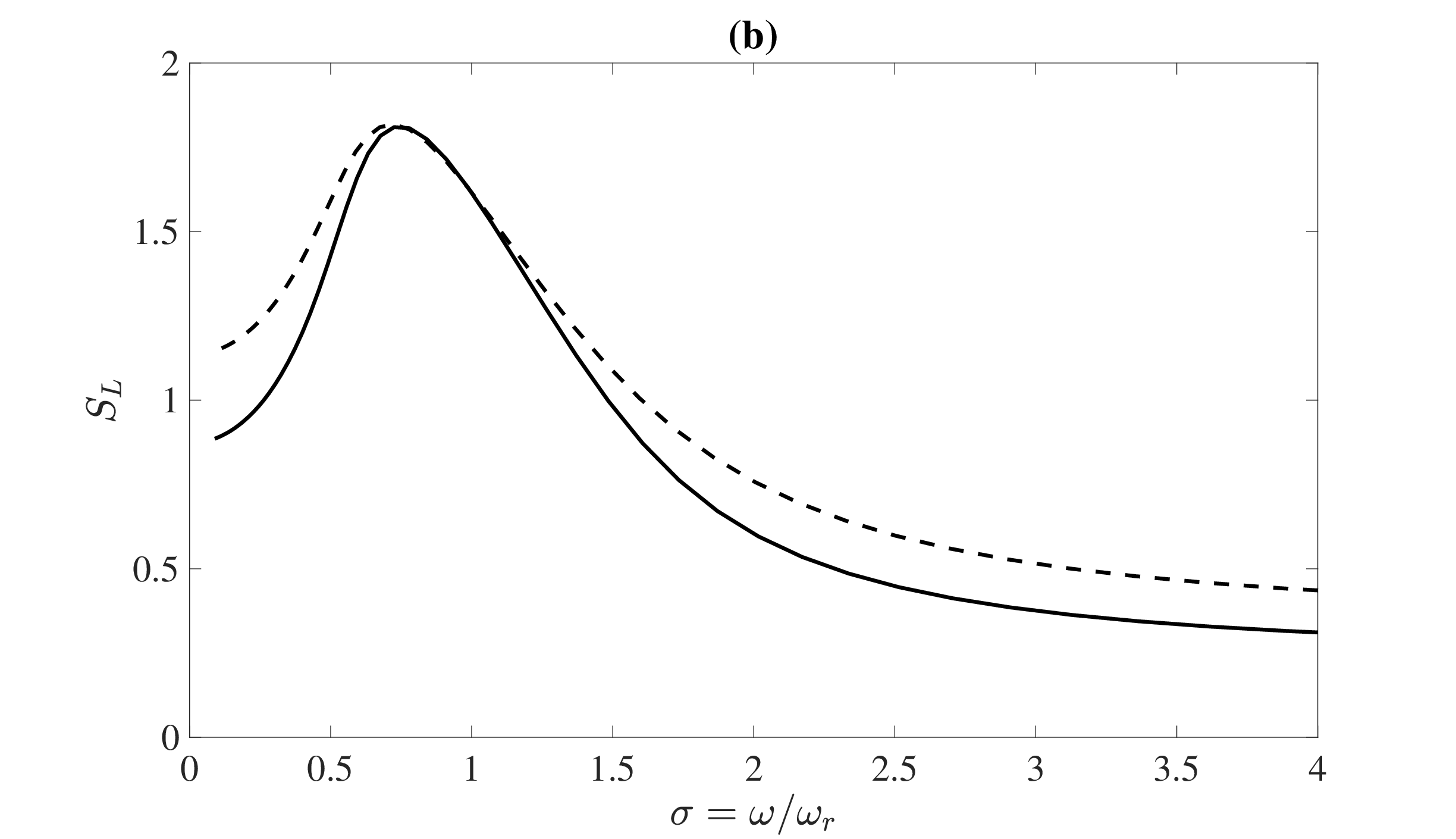,width=.55\linewidth}}
  \caption{Stride length $S_L$ vs. $\hk_a$ (a) and vs.  $\sigma$ (b) for a rigid foil with prescribed heave ($h_0=0.2$) computed using Garrick's thrust (dashed lines) and the present $\oC_T$ (solid lines), for $R=0.48$, $\hb_a=0$ and $\oC_D=0.25$.}
\label{fig_Gra}
\end{figure}

For the self-propulsion problem, $k$ is not given but obtained form \eqref{CDCT} by the iterative process described at the end of \S \ref{sec_sol_pro}, as done by \cite{grapa26}, who used Garrick's \citeyear{garri36} expression for the time-averaged thrust $\oC_T$ of a rigid foil.  Here the results are obtained with the  expression of $\oC_T$ for the more general flexible foil commented on in section \ref{sec_for}. With either expression for $\oC_T$, the solution for $k$ (or $S_L$) will depend on $R$, $\hk_a$,  $\hb_a$, $\oC_D$, and $h_0$, in this  particular case of  a rigid foil with prescribed heave. 

To have an idea of the differences when the two thrust  formulations are used, Fig. \ref{fig_Gra} shows the stride length $S_L$ versus $\hk_a$, and also versus $\sigma=\omega/\omega_r$, for the same set  of parameters used  in Fig. 5 of \cite{grapa26} (in the present notation, $R=0.48$, $\hb_a=0$, $\oC_D=0.25$ and $h_0=0.2$). The dashed curve in Fig. \ref{fig_Gra}(b)  coincides with Fig.  5 in \cite{grapa26} since it is obtained with the same formulation (same expression for $\oC_T$ and same Eq. \eqref{alpharig2}; though, alternatively, we can solve the system \eqref{sistema} with large $\hS$ and $\hk_h$ to fix $d_{1}=d_{2}=0$ and $h_0=0.2$, and $\omega_r$ by minimizing $|\det(\tA)|$, which yield exactly  the same results as indicated above). 
The stride length from the present model for $\oC_T$ (continuous line) is practically the same for the maximum cruising velocity, that occurs near the resonance frequency of the torsional spring, but differs slightly as the frequency departs from it. It must be noted that the present $\oC_T$  for a rigid foil has shown a  better general agreement with  numerical and experimental results than Garrick's model for the time-averaged thrust coefficient \citep[e.g.,][]{ferna17,alami21}, while for a flexible foil 
has shown a reasonable agreement with available small-amplitude experimental data \citep{ferna25}. 

The representations of $S_L$ (or any other non-dimensional quantity of interest) vs. $\hk_a$, or $k_a$, or $\omega/\omega_r$,  are very useful because they are universal (for a given set of the rest of non-dimensional parameters)  and  provides simple rules for the tuning of dimensional quantities such as frequency, spring stiffness $\tilde{k}_a$ or cruising velocity $U$ to attain maximum stride length or maximum efficiency.  Although the representation versus $\omega/\omega_r$ (Fig. \ref{fig_Gra}(b)) is more `visual' physically than the representation versus $\hk_a$ (Fig. \ref{fig_Gra}(a)), because the former tell us how close is the optimal performance to the resonant state of the mechanical system,  the last one  is arguably more rational because $\hk_a$ is one of the independent non-dimensional parameters governing the problem, whereas the natural frequency $\omega_r$ has to be computed for the given  set of these parameters ($\omega/\omega_r$ can also be represented vs. $\hk_a$ for a given set of the rest of non-dimensional parameters). As we shall see below (e.g., \S \ref{sec_res_rig}), this difficulty is more apparent when results are depicted in terms of more than one dimensionless parameter simultaneously. 

\begin{figure}
\centerline{\epsfig{file=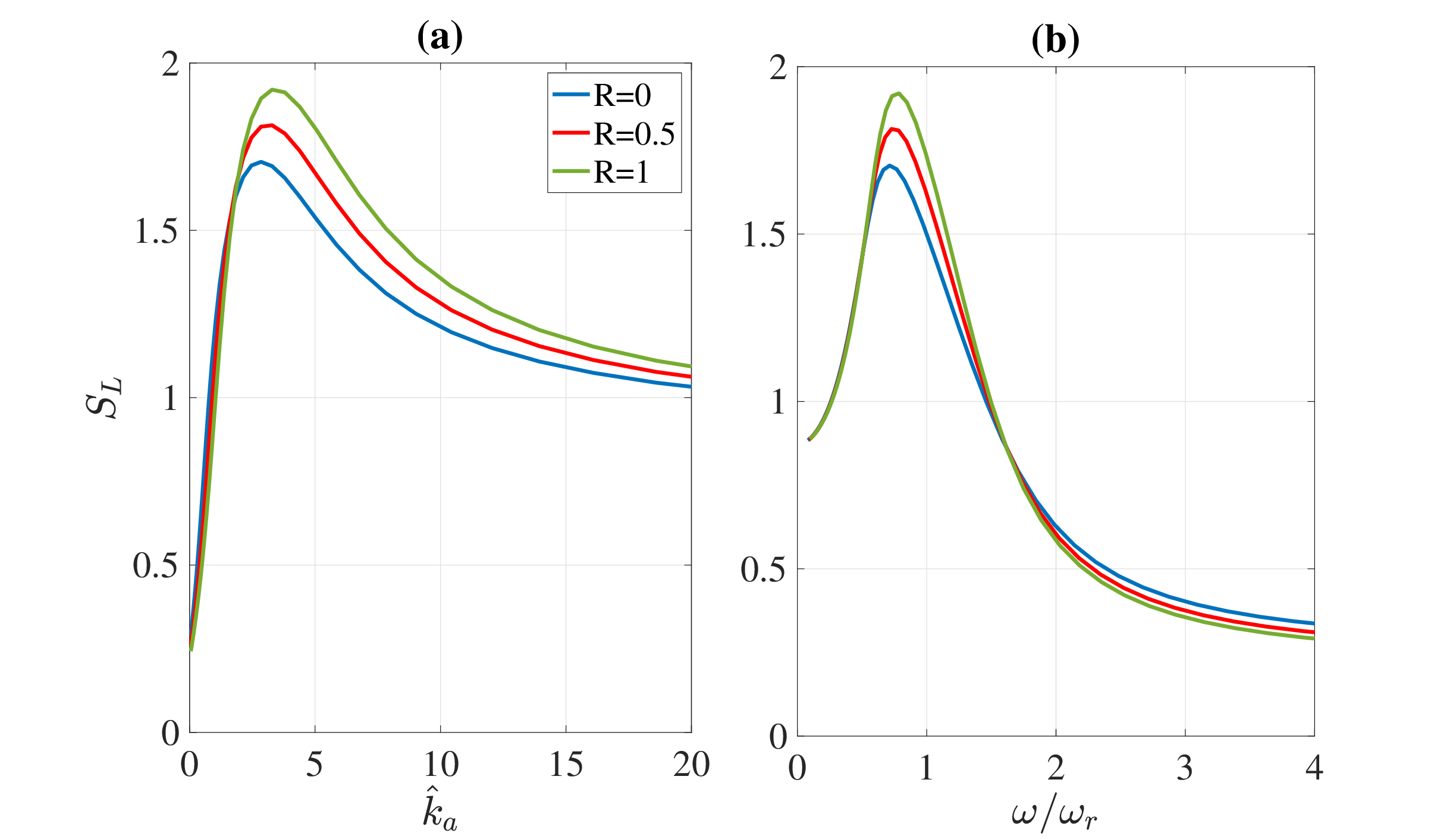,width=.55\linewidth}\epsfig{file=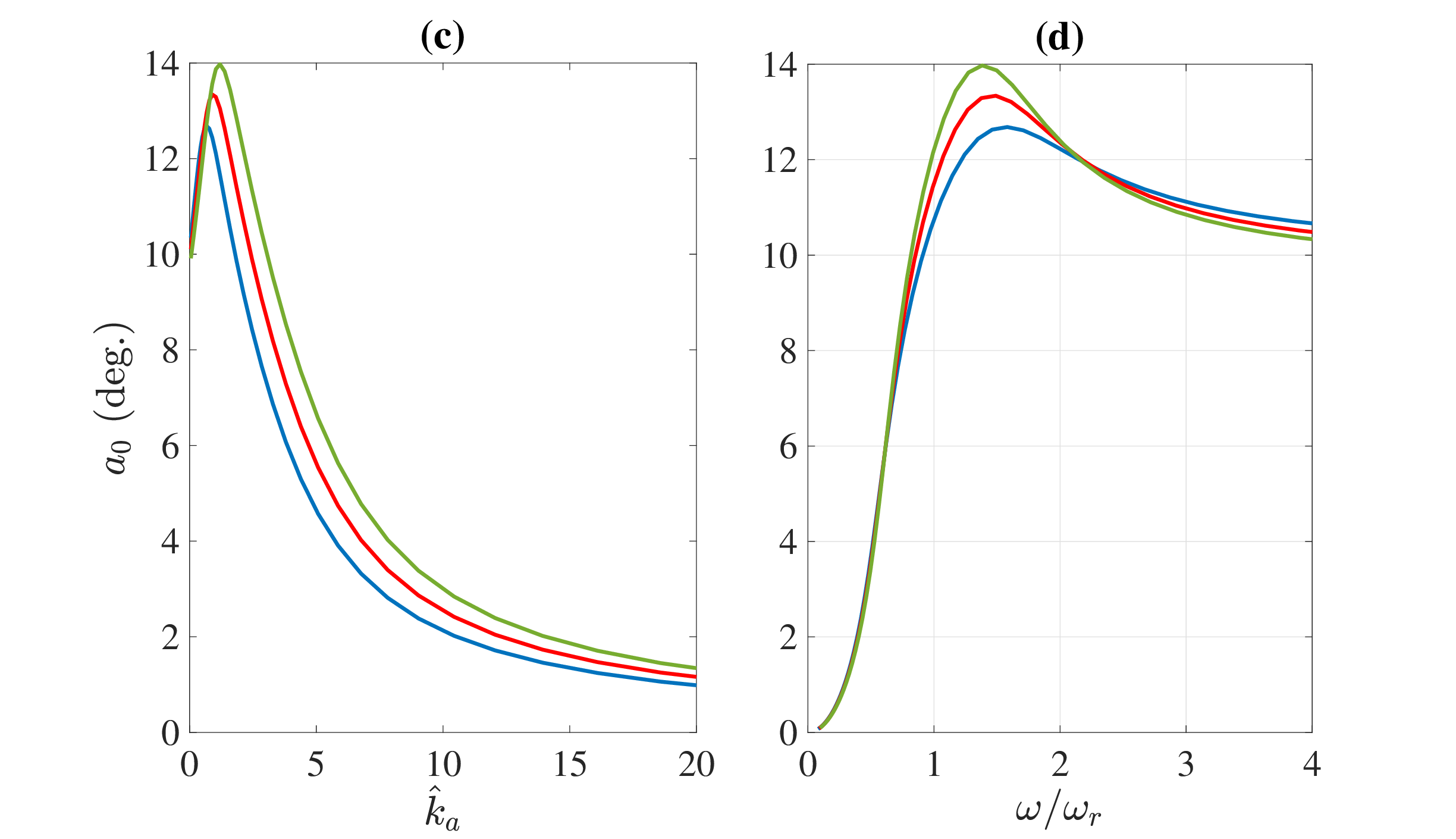,width=.55\linewidth}}
\centerline{\epsfig{file=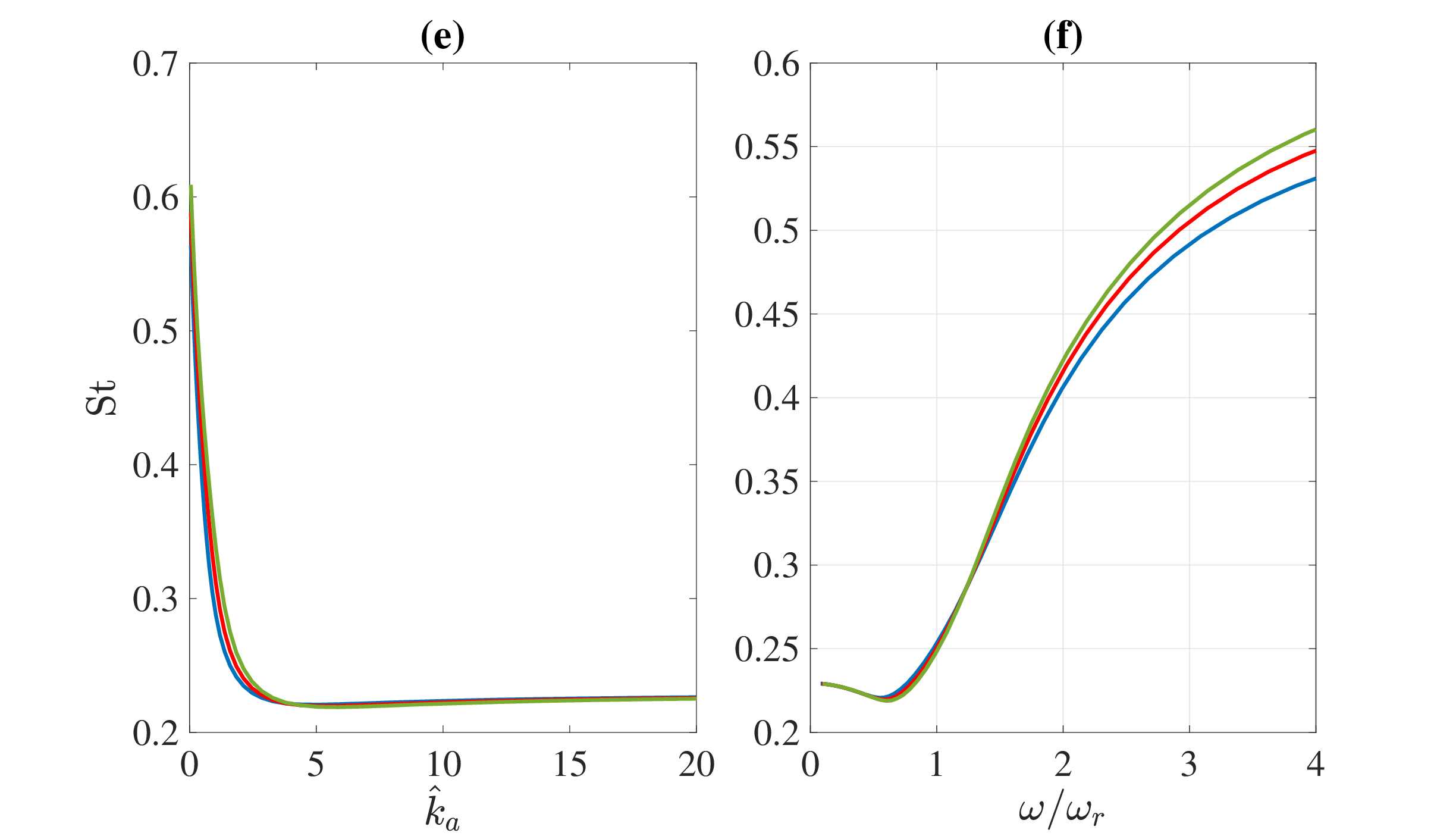,width=.55\linewidth}\epsfig{file=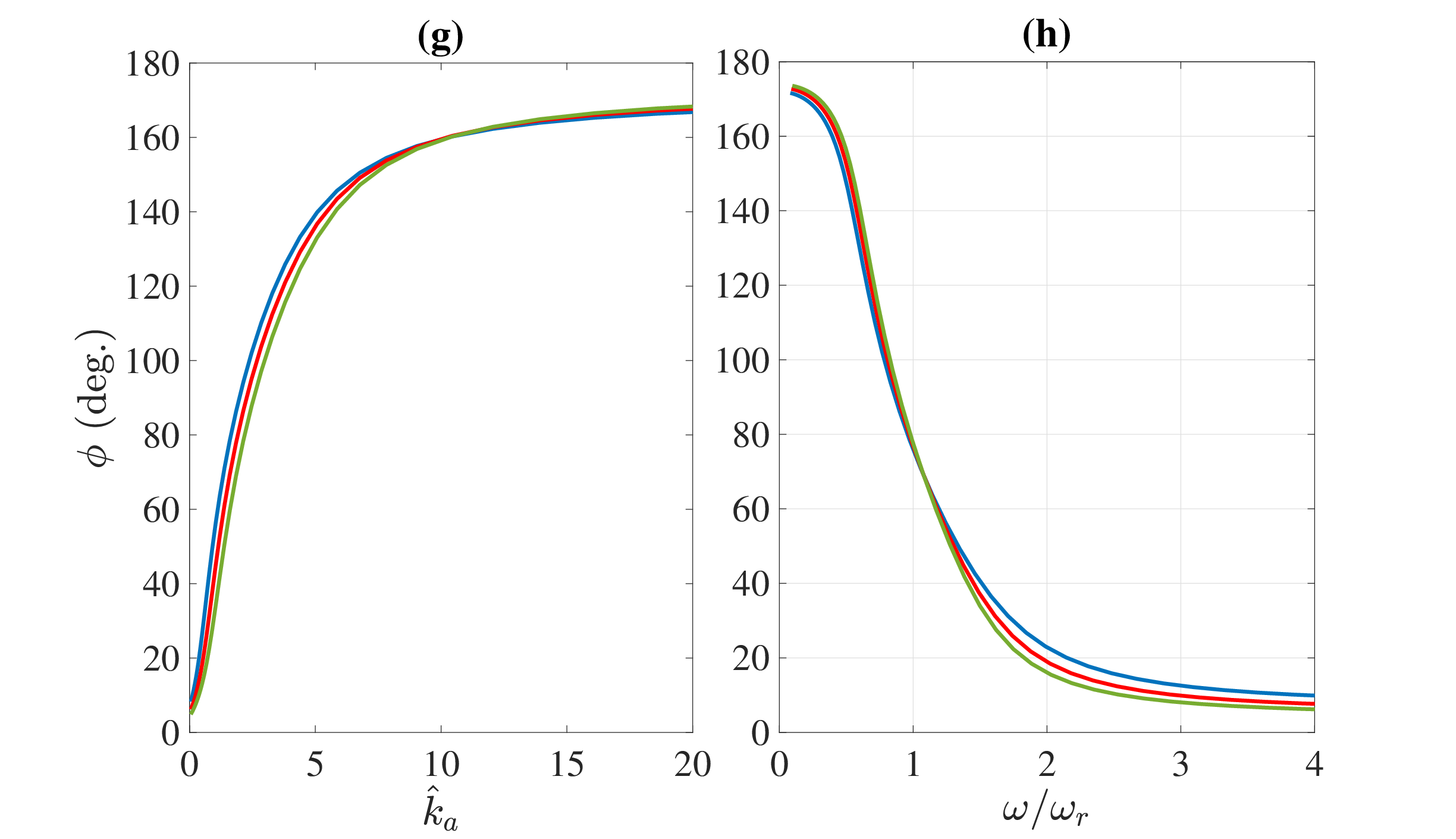,width=.55\linewidth}}
  \caption{Stride length (a-b), pitch angle (c-d), Strouhal number (e-f) and pitch/heave phase shift (g-h) as a function of $\hk_a$ and $\omega/\omega_r$ for the same case considered in Fig. \ref{fig_Gra}, but for tree values of the mass ratio, as indicated in (a).}
\label{fig_rigid1}
\end{figure}

To appreciate how the results are affected by the mass ratio $R$  in both representations for this particular case considered by \cite{grapa26} with $\hb_a=0$, $\oC_D=0.25$ and $h_0=0.2$, but with the present $\oC_T$, Fig. \ref{fig_rigid1} shows the stride length $S_L$, the pitch angle and phase, $a_0$ and $\phi$, and the Strouhal number St for $R=0$, $R=0.5$ and $R=1$, thus covering more than the whole range of mass ratios of interest in aquatic locomotion \citep{floro18}. The variation of all the plotted properties with $R$ is rather small. The peaks of the stride length and the pitch angle grow with $R$, appearing at slightly increasing values of $\hk_a$, but corresponding to practically the same value of $\omega/\omega_r$, which is about $0.75$ for $S_L$ and  $1.4$ for $a_0$ (Figs. \ref{fig_rigid1}b and  \ref{fig_rigid1}d, respectively). The maxima of $S_L$ as $R$ is varied correspond to practically the minima of the Strouhal number, about $0.22$ (Fig. \ref{fig_rigid1}f), which is within the range  experimentally reported for optimal aquatic locomotion \citep{taynu03}, and to a phase shift $\phi$ of about $90^\circ$, consistent with previous results for optimal  propulsion using pitching and heaving rigid  foils \citep{andst98,quila15}.

\begin{figure}
\centerline{\epsfig{file=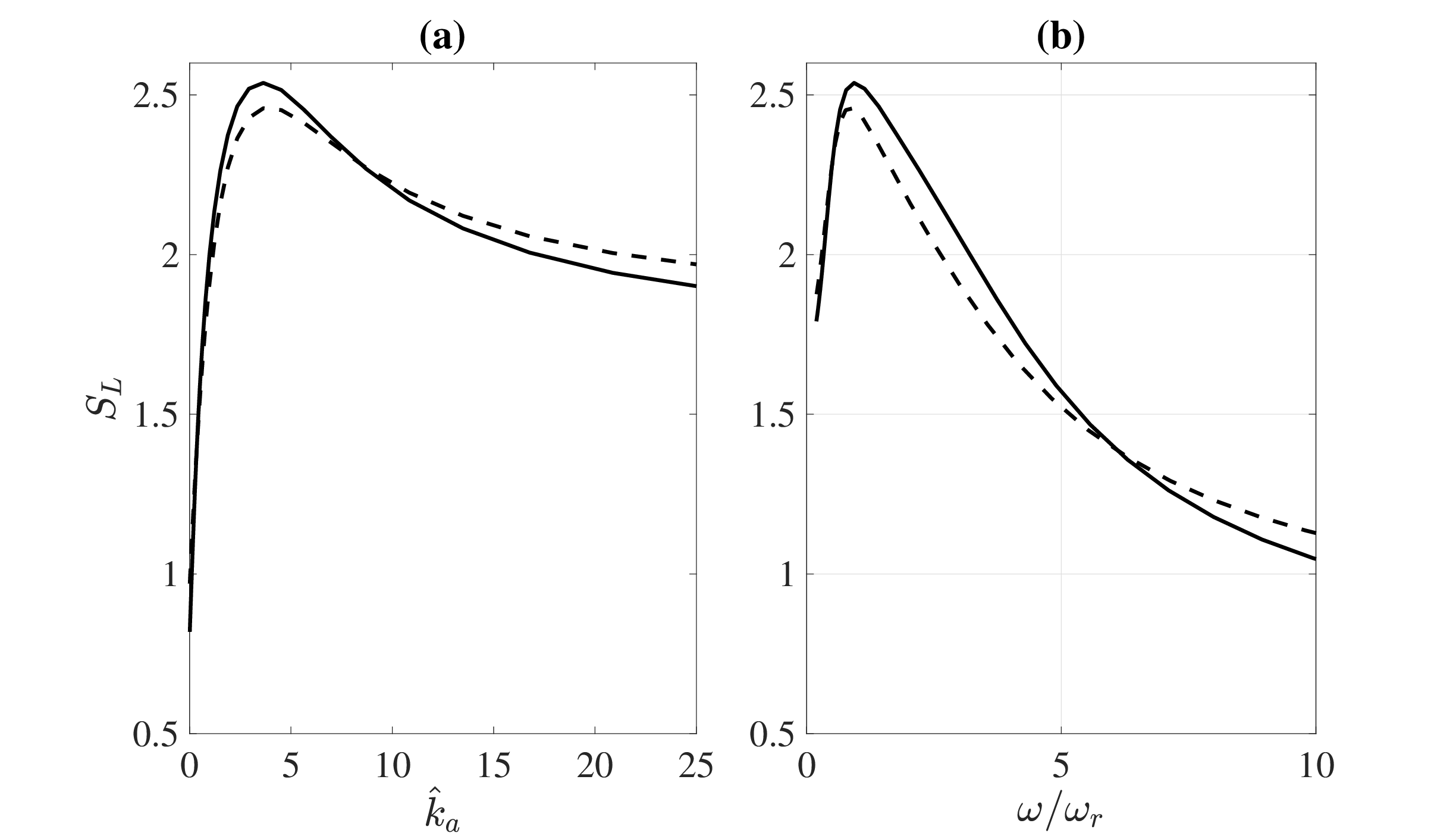,width=.55\linewidth}\epsfig{file=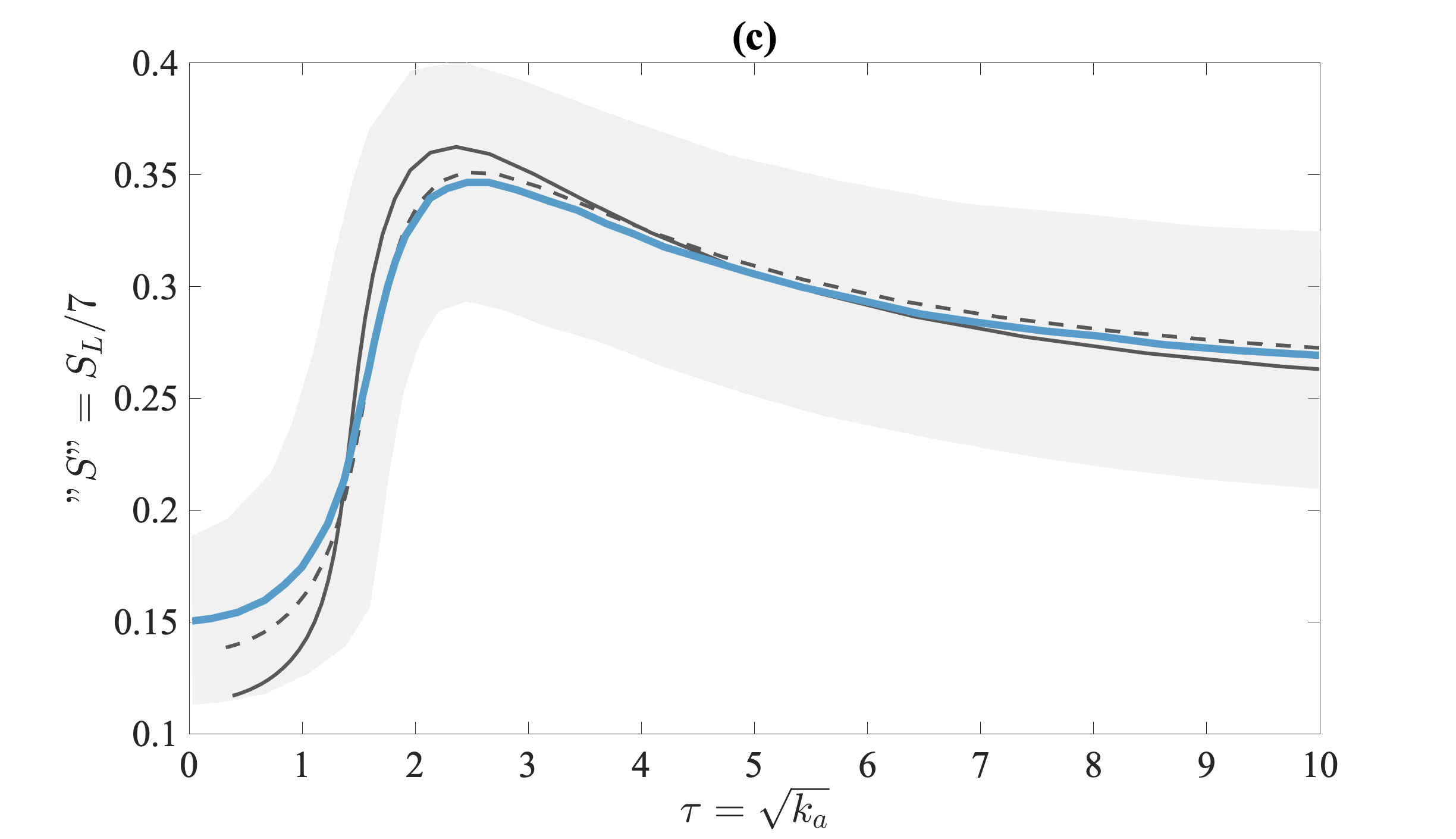,width=.55\linewidth}}
\caption{Stride length $S_L$ vs. $\hk_a$ (a) and vs. $\omega/\omega_r$ (b) for the case considered in the model by \cite{zhozh21} (see data in the main text) computed with Garrick's $\oC_T$ (dashed lines) and the present model for $\oC_T$ (solid lines).  (c): Same results but plotted as the stride length in body lengths $"S"$ vs. the scaled muscle/spring tension $\tau$; the blue line is the result from the model developed by \cite{zhozh21}, taken from their Fig. 3A together with the error band for their experimental results.}
\label{fig_zhong}
\end{figure}

The model developed by \cite{grapa26} is quite similar to the locomotion model of fish-like robots previously considered by \cite{zhozh21}, with a rigid tail joined to the body of the robot through a peduncle with a tunable stiffness in such a way that  when a prescribed heaving motion is transmitted to the tail,  its passive pitching movement, which depends on the rigidity of the tail joint, can be tuned to generate optimal thrust. \cite{zhozh21} compared with great success the results of their simple theoretical model that  tunes  tail stiffness using a motor-driven `muscle' (spring)  with experimental measurements using a robot prototype that simulates a tuna swimming (see their Fig. 3A and the present Fig. \ref{fig_zhong}c commented on below). The model is summarized in Eq. (1) of \cite{zhozh21} for the displacement of the tail fin's trailing edge, modelling the elastic joint through a torsional spring of variable stiffness, like in the present and previous similar models. This expression (1) comes, in the present notation,  from equation \eqref{alpharig1}  for $\alpha(t)$ taking into account that the trailing edge (dimensional) amplitude for a rigid foil  is $c z_s(x=1,t)= c(h-2 \alpha)$, with $h(t)$ prescribed. Actually, the equivalent expression to \eqref{alpharig1} for $\alpha(t)$ is given by Eq. 2.5 in S2 of the supplementary materials of \cite{zhozh21}. Taking into account that their dimensionless time is $T=2 \pi f \tilde{t}$,  instead of the $t=2U \tilde{t}/c$ employed here, so that $d / d T=(S_L/\pi) (d / d t)$, this Eq. 2.5 is Eq. \eqref{alpharig1} if $R=0$, $b_a=0$, assuming a two-dimensional tail with $c=l_F$ and $L_T\equiv l_T/l_F \to \infty$, and one  identifies their scaled muscle/spring tension $\tau$ with the square root of the present dimensionless stiffness of the torsional spring $k_a$, both appearing in the same factor $(\pi \C(k)/2+ k_a)$ multiplying the $\alpha-$term in their Eq. 2.5 and in the present \eqref{alpharig1}; i.e., 
\be \tau^2 = k_a \,. \label{tau_ka}
\ee
Note that this non-dimensionless spring stiffness $k_a$  is the one scaled with $U$ in Eq. \eqref{springs}, not the $\hk_a$ scaled with $\omega$ in Eq. \eqref{paramo2}. That's why we have written Eq. \eqref{alpharig1} with $k_a$ instead of using the more appropriate \em hatted \rm parameters for the self-propulsion problem, employed in Eq. \eqref{alpharig2} and in all the  reported computations of the present work, though the results can afterwards be given in terms of $k_a$ once $k$ (i.e., the cruising speed $U$) is computed . The use of the spring stiffness non-dimensionalized with the cruising speed $U$ is relevant in the work by \cite{zhozh21} because their main conclusion is that the optimal locomotion with maximum stride length is attained by tuning the (dimensional) spring tension $\tilde{k}_a$ with the swimming speed squared, so that $k_a$ remains constant.

To reproduce the results by \cite{zhozh21} with the present model, in addition to the equivalence \eqref{tau_ka}, we take into account that $l_F=c=30$ mm and the body length, which is used to scale $\oC_T$ and define the stride length, is $l=350$ mm, so that $\oC_D = 0.046 (l/l_F)^2=2.254$ and their stride length, defined as body lengths traveled per tailbeat, is $"S"=(l_F/l)S_L=S_L/7$. On the other hand, also from S2.1 in the supplementary materials, the non-dimensional heave amplitude and mass ratio are,  in the present notation,   $h_0=2 \times 0.64=1.28$ and $R=4\times 0.01=0.04$, respectively (though they use $R=0$ in their model, as indicated above). With this set of parameters we compute $S_L$ using both Garrick's model and the present one for $\oC_T$, and plot them in Figs. \ref{fig_zhong}a and \ref{fig_zhong}b in terms of both $\hk_a$, $\omega/\omega_r$, respectively. Figure \ref{fig_zhong}c shows the same stride lengths, but in body lengths, $"S"=(l_F/l)S_L$, versus $\tau=\sqrt{k_a}$, comparing them with the results from Zhong et al.'s model depicted in their Fig. 3A (blue line in Fig. \ref{fig_zhong}c). The minor discrepancies with the results from the present model using Garrick's thrust (dashed line) are due to the possible slight inaccuracies in the conversion factors. The three curves present a maximum of the stride length at $\tau \simeq  2.5$, corresponding in Figs. \ref{fig_zhong}a and to $\hk_a \simeq 3.5$, and $\omega/\omega_r$ slightly smaller than unity in Fig.  \ref{fig_zhong}b.  It should be noted that the use of $k_a$ to represent the results is not, \em a priori\rm, so appropriate as the use of $\hk_a$ because the former scales the dimensional spring constant $\tilde{k}_a$ with the square of the unknown cruising speed $U$ whereas  the later uses the given actuating frequency $\omega$, and therefore it is a given known parameter (see \S \ref{sec_for}). Actually, depending on the values of the remaining parameters, specially $h_0$ and $\oC_D$, it is quite common that the resulting dependence of $S_T$ with $k_a$ is multi-evaluated,  because so it is the dependence of $k_a$ with $\hk_a$, though in the present particular case it is not.

Also shown in Fig. \ref{fig_zhong}c is the band error for the experimental results by \cite{zhozh21} from their Fig. 3A.  The comparison indicates that the present simple linearized  model  is, like theirs,  an excellent approximation to analyze the locomotion performance of this kind of fish-like robots, even for values of the heave amplitude as high as $h_0=1.28$! The present model has the advantage, in relation to these previous ones, -- independently of,  and  in addition to, a presumably better accuracy of the present $\oC_T$ model for  a wider range of frequencies,  as indicated above --  that it is extended to include the effect of passive heave in aquatic locomotion, to be considered next for a rigid foil (\S \ref{sec_res_rig}). And,  what it is more relevant, it includes the effect of the foil flexibility down to non-dimensional stiffnesses of order $10^{-1}$ (up to the second natural frequency of the fluid-foil system). Therefore, both the flexibility of the foil and the rigidity of its elastic joints can be jointly tuned to refine the search for optimal locomotion
 (see \S \ref{sec_res_fle}).

\begin{figure}
\centerline{\epsfig{file=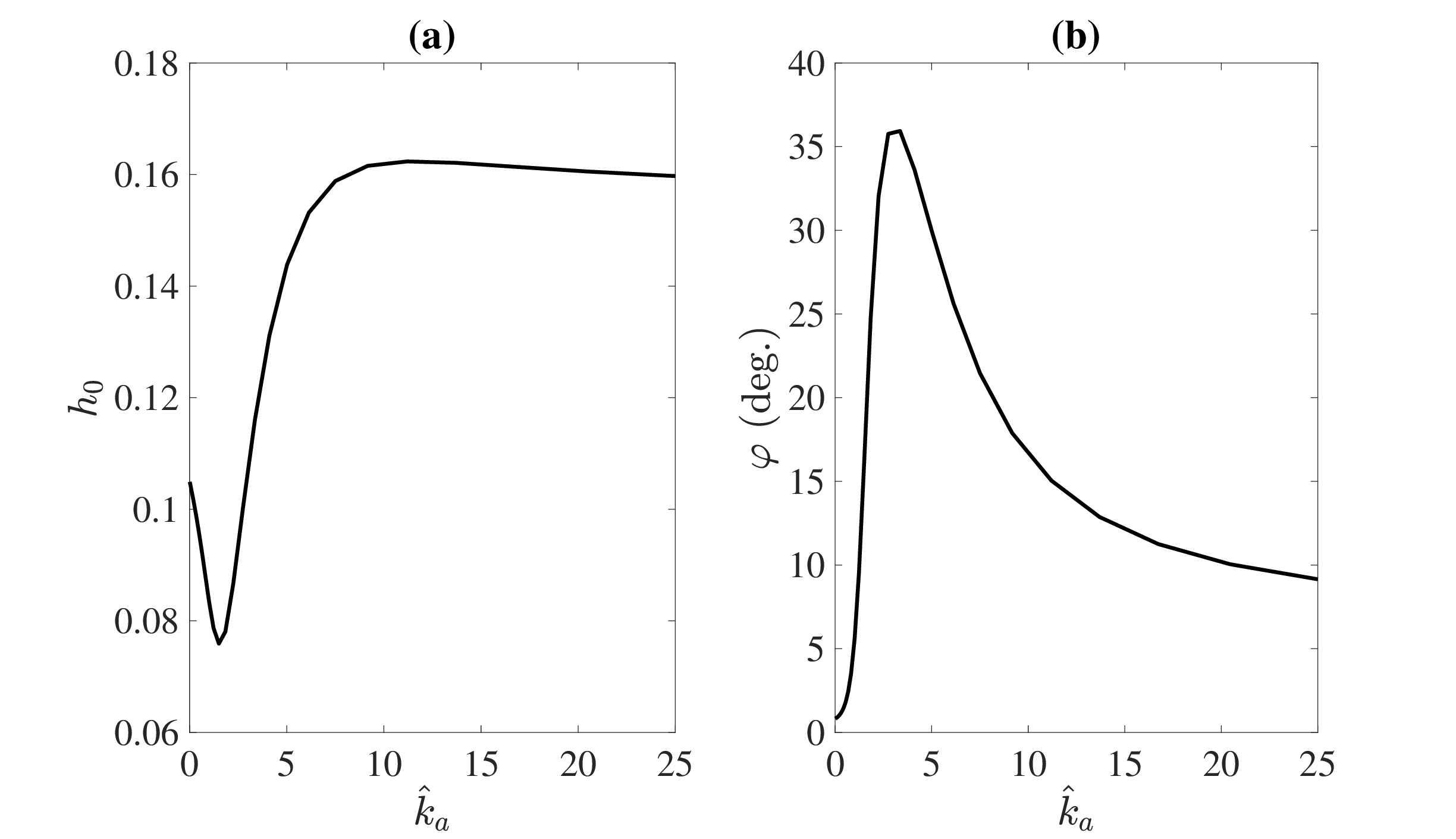,width=.55\linewidth}\epsfig{file=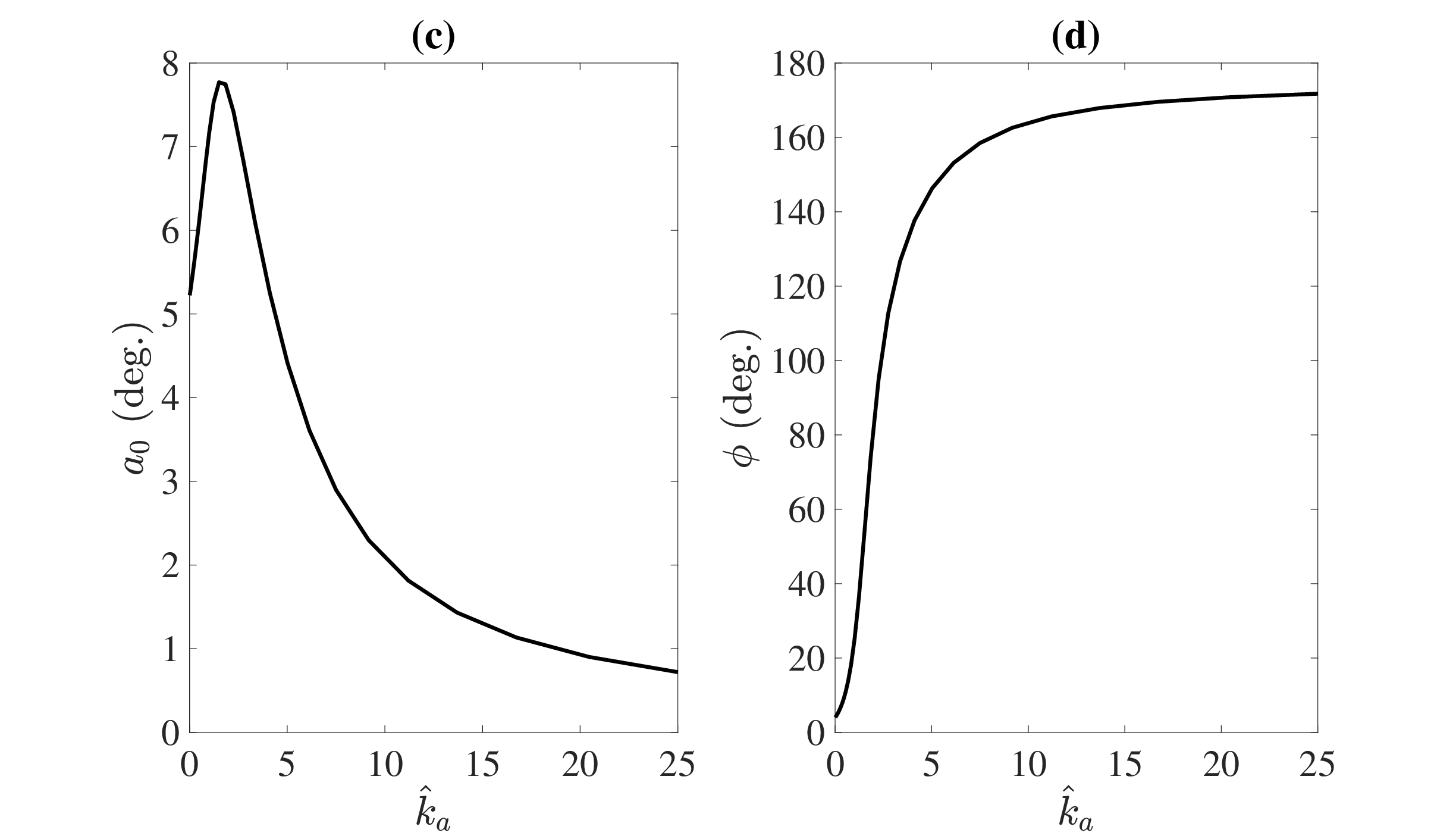,width=.55\linewidth}}
\centerline{\epsfig{file=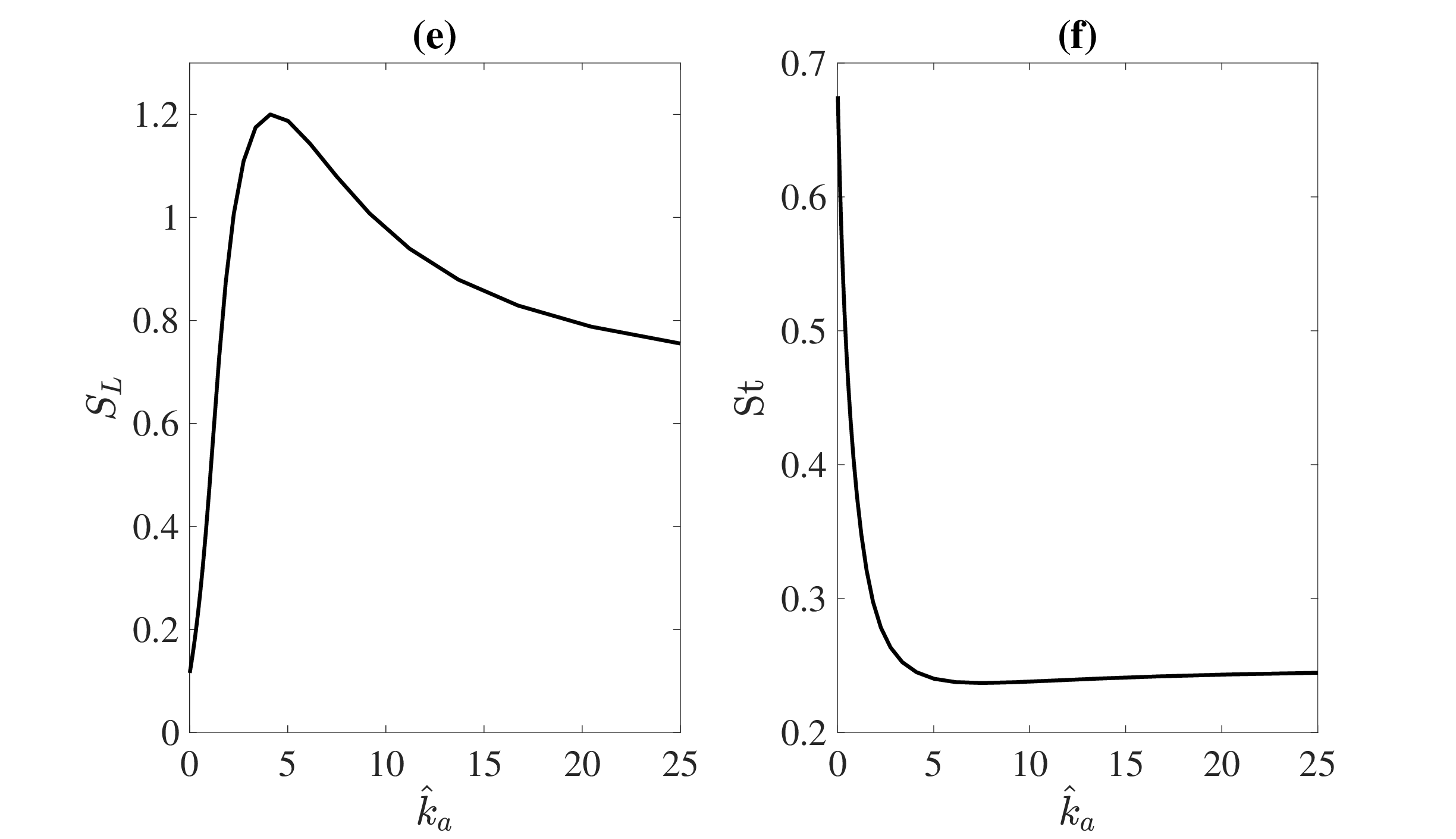,width=.55\linewidth}\epsfig{file=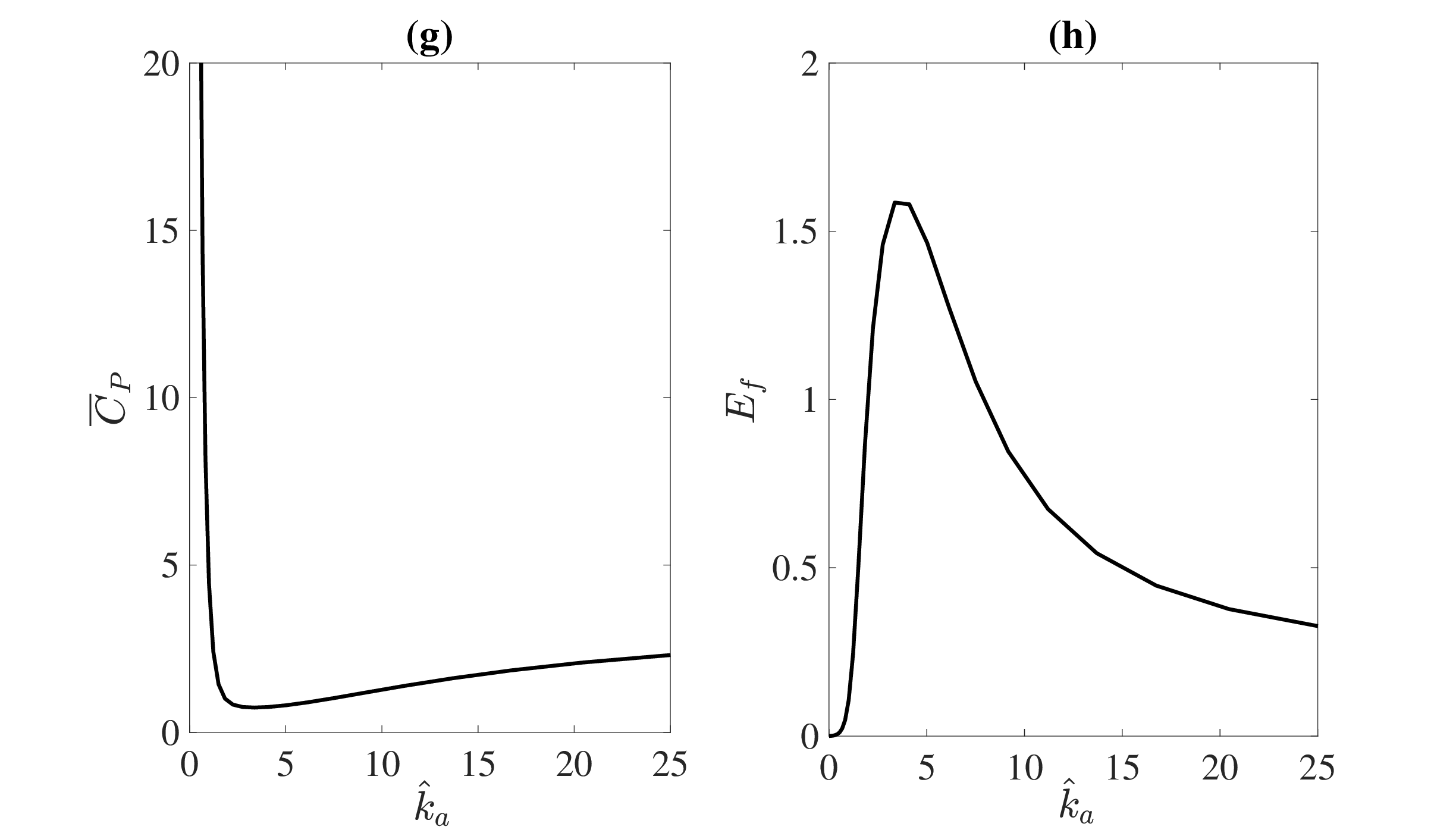,width=.55\linewidth}}
  \caption{Heave amplitude (a) and phase (b), pitch amplitude (c) and phase (d), stride length (e), Strouhal number (f), power input (g) and efficiency (h) vs. $\hk_a$ for a rigid foil ($\hat{S} \to \infty$) with $R=0.48$,  $\hk_h=10$, $\hb_a=\hb_h=0.05$, and  for $\hA_l =1$ and $\oC_D=0.25$.}
\label{fig_rigid2}
\end{figure}

\subsection{Results for a rigid foil with passive heave and pitch}
\label{sec_res_rig}

Figure \ref{fig_rigid2} shows some representative results for a foil actuated by a force with non-dimensional  amplitude $\hA_l =1$ as a function of the torsional spring stiffness $\hk_a$,  for a given set of the remaining parameters. Now, the heave amplitude $h_0$ (Fig. \ref{fig_rigid2}a) is not constant, but vary according to the applied force, the fluid forces and moments, the inertia of the foil and the stiffnesses of the torsional and linear springs. Since a relatively high value of the last one has been selected ($\hk_h=10$), the heave amplitude vary  around $0.1$ for the given force amplitude $\hA_l=1$ ($h_0\simeq \hA_l/\hk_h$), with a minimum  at $\hk_a \simeq 1.5$, where  the pitch amplitude has a pronounced peak (Fig. \ref{fig_rigid2}c). The stride length and the efficiency have their maxima practically at the same value $\hk_a \approx 4$ (Figs. \ref{fig_rigid2}e and \ref{fig_rigid2}h), where, approximately, the Strouhal number has its minimum value (Fig. \ref{fig_rigid2}f) and the pitch phase shift is about $90^\circ$ (Fig. \ref{fig_rigid2}d). The variation of the heave phase (Fig. \ref{fig_rigid2}b) is more moderate than the pitch phase, and presents a pronounced peak at $\hk_a \simeq 3$. It must be recalled  that $\varphi$ is actually the phase shift of the input force in relation to the heave, so that all the phase shifts take the heaving motion as a reference.

The pattern shown in Figure \ref{fig_rigid2} for the different locomotion parameters  as the torsional spring stiffness $\hk_a$ is varied remains quite similar when the input force $\hA_l$ changes, provided that  $\hk_h$ is sufficiently high (see below). Obviously, as $\hA_l$ increases, both $h_0$ and $a_0$ grows almost linearly because the linearity of Eq. \eqref{sistema} (the amplitudes  do not increase exactly proportional to $\hA_l$  because the scalar terms in the unknown vector ${\bf{d}}_0$ are complex quantities, and their phases vary jointly with their amplitudes as $\hA_l$ changes). Consequently, the peak value of the stride length  grow almost linearly with the applied force, though it is found that the corresponding Strouhal number remains almost constant. Since the present linear formulation is valid for small values of $h_0$ and $\alpha_0$, the solution eventually fails as  $\hA_l$ increases. For this reason, most of the reported results will be obtained using $\hA_l=1$ (or  smaller), which, according to Fig. \ref{fig_rigid2}, provides results for $h_0$ and $a_0$ within the validity range of the theory. In any case, since $\hA_l$ is the physical input force amplitude $L_{i0}$ non-dimensionalized (see Eq. \eqref{inlift2}), in practice one can always select the  frequency and/or the chord length to fix $\hA_l=1$ for a given $L_{i0}$, or vice-versa.

\begin{figure}
\centerline{\epsfig{file=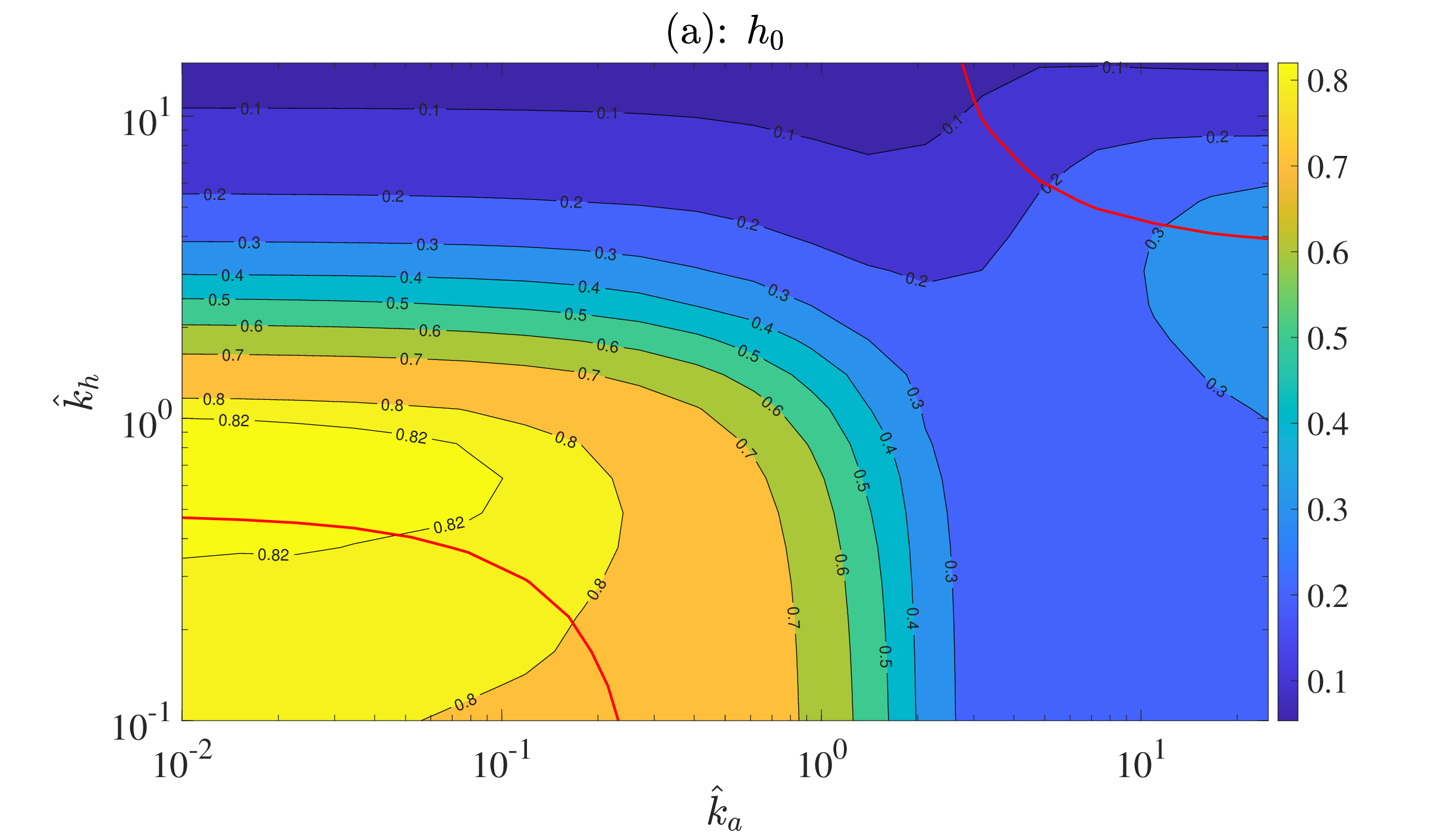,width=.55\linewidth}\epsfig{file=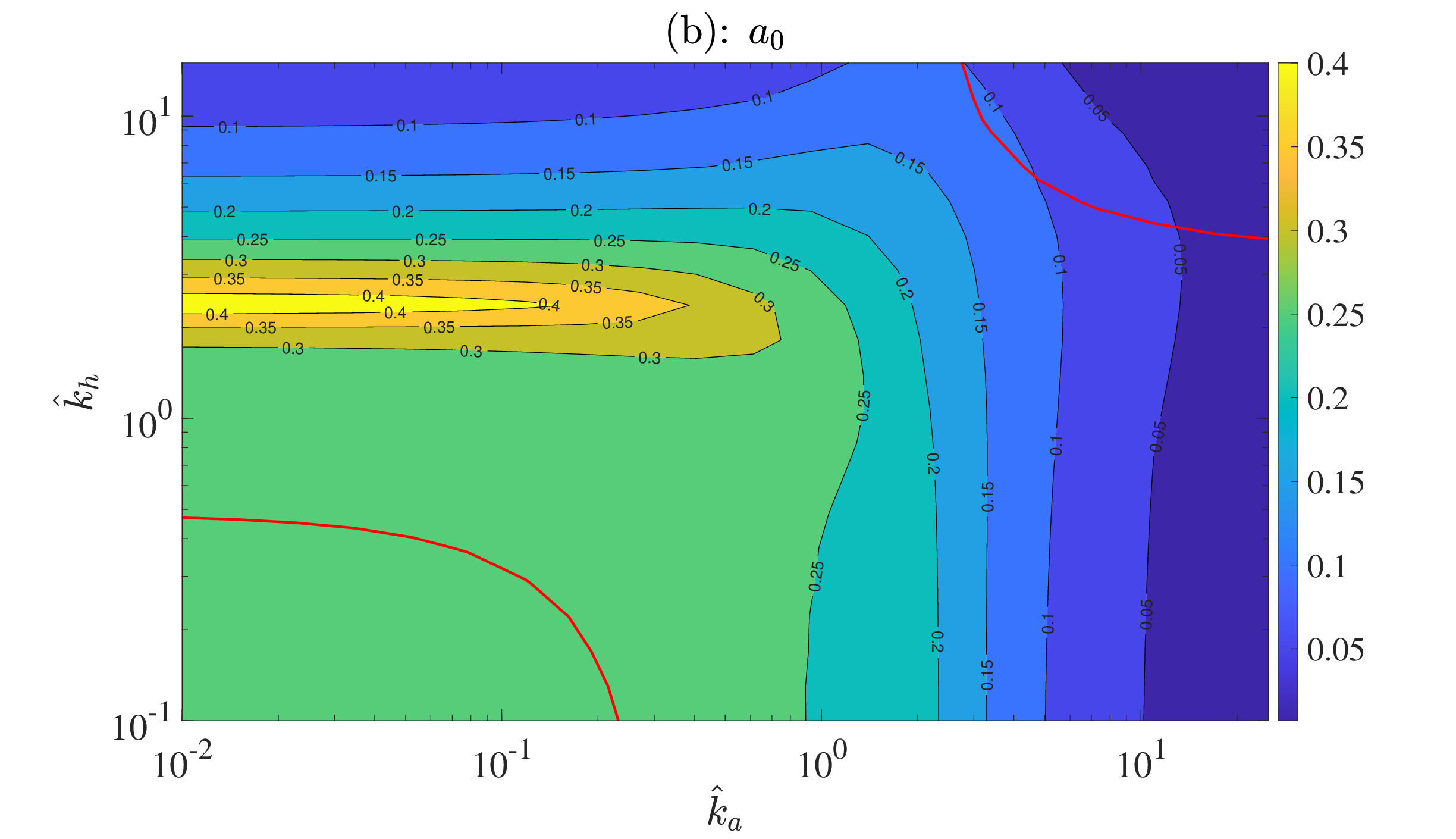,width=.55\linewidth}}
\centerline{\epsfig{file=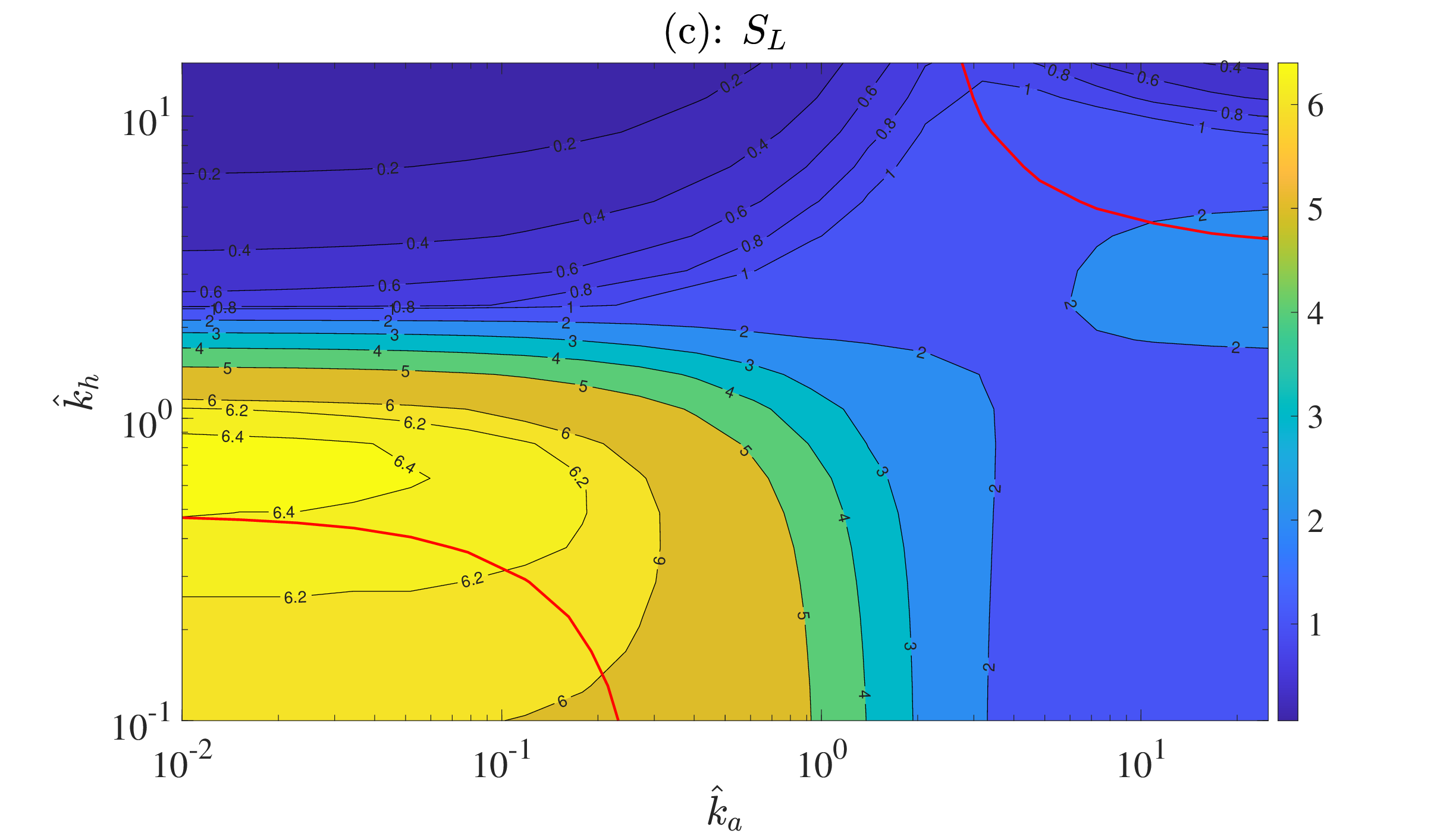,width=.55\linewidth}\epsfig{file=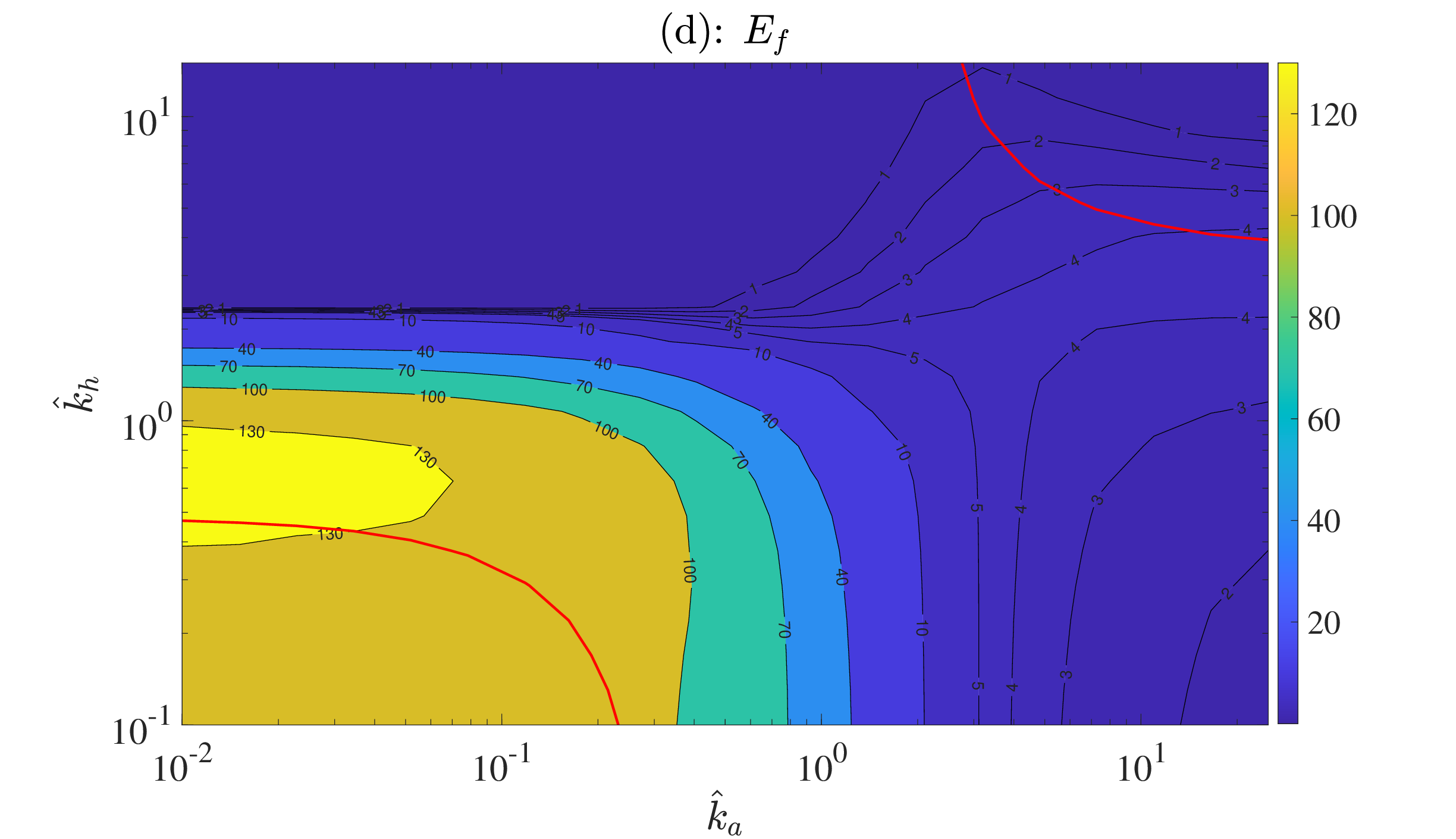,width=.55\linewidth}}
  \caption{Heave (a) and pitch (b) amplitudes, stride length (c),  and efficiency (d) in the $(\hk_a,\hk_h)-$plane for a rigid foil ($\hat{S} \to \infty$) with $R=0.48$,  $\hb_a=\hb_h=0.05$, and  for $\hA_l =1$ and $\oC_D=0.25$. The red lines correspond to the two branches of the resonant frequencies approximation \eqref{kr0ah}.}
\label{fig_rigid3}
\end{figure}

To see the effect of the rigidity $\hk_h$ of the translational spring, jointly with the effect of  $\hk_a$, Fig. \ref{fig_rigid3} shows results in the $(\hk_a,\hk_h)-$plane for the same values of the remaining parameters as in Fig.   \ref{fig_rigid2}. It is observed that the stride length and efficiency peaks observed in Fig.   \ref{fig_rigid2}  for $\hk_h=10$ (occurring at $\hk_a \approx 4$) are  local maxima, but much smaller than their maxima when $\hk_h$ is slightly below unity and $\hk_a$ is small (Figs. \ref{fig_rigid3}c and \ref{fig_rigid3}d). 

Both local maxima are associated to the springs (torsional and translational) natural frequency, which has two branches in the $(\hk_a,\hk_h)-$plane that can be obtained 
by minimizing $|\det(\tA)|$ to compute the non-dimensional resonant frequency $k_r$ and setting \eqref{sigr} to unity. But, instead of that, it is better to use an analytical approximation of $k_r$ for a rigid foil, 

\small
\be k_{r0}=\sqrt{\frac{\pm \sqrt{[48 k_a (R+\pi )+k_h (32 R+27 \pi )]^2-192 k_a k_h (R+\pi ) (8 R+3 \pi )}+48 k_a (R+\pi )+k_h (32 R+27 \pi )}{2(R+\pi ) (8 R+3 \pi )}} \,,  \label{kr0ah} \ee
\normalsize
which is used in Fig. \ref{fig_rigid3} to represent $\omega/\omega_{r0}=k/k_{r0}=1$  (red lines). These are mixed torsional and translational springs natural frequencies, with a translational spring branch corresponding to the positive sign (lower curve in the figure) and the torsional spring branch corresponding  to the negative sign (upper curve). The expression \eqref{kr0ah} contains the inertial contributions associated to $\tA_0$ given by Eq. \eqref{def_A0} (terms with $R$) and part of the FSI contributions (terms with`$\pi$'). These last ones come from the large-$k$ approximation of $\tA_f$ given by Eq. \eqref{Af}. That is,  retaining only the $k^2$ terms in Appendix \ref{app_coeffA} to solve $|\det(\tA)|=0$ (for $S\to \infty$ and $b_a=b_h=0$). Although they are mixed modes,  to simplify the branch with positive sign will named translational spring mode and that with negative sign torsional spring mode. In fact, without considering the FSI, for $k_a \to \infty$ and $k_h \to \infty$ one obtains the well-known results $k_{r0} \to \sqrt{k_h/R}$ and $k_{r0} \to \sqrt{3 k_a/2 R}$,  with the plus and minus signs, respectively. 

The largest values of $S_L$ and $E_f$ are associated to a pronounced peak of the heave amplitude for $\hk_h \approx 0.65$ (see Fig. \ref{fig_rigid3}a). This maximum of $h_0$ is  much larger than the local one for $\hk_h=10$, also observed in Fig. \ref{fig_rigid3}a, and which is associated to the  torsional spring mode  of the springs resonance.  Note in Fig. \ref{fig_rigid3}b that this  resonance generates a maximum pitch amplitude at $\hk_h \approx 2.4$ for small $\hk_a$, but it extends as $\hk_a$ increases towards larger values of $\hk_h$, so that the peak pitch amplitude for $\hk_h=10$ is the one plotted in Fig. \ref{fig_rigid2}c at about $\hk_a=1.5$. When the heave is prescribed instead of passive, like in the models reported in the previous subsection \ref{sec_pre_rig}, only the torsional spring mode is present and one cannot achieve the higher stride length and efficiency associated to the translational spring mode.

\begin{figure}
\centerline{\epsfig{file=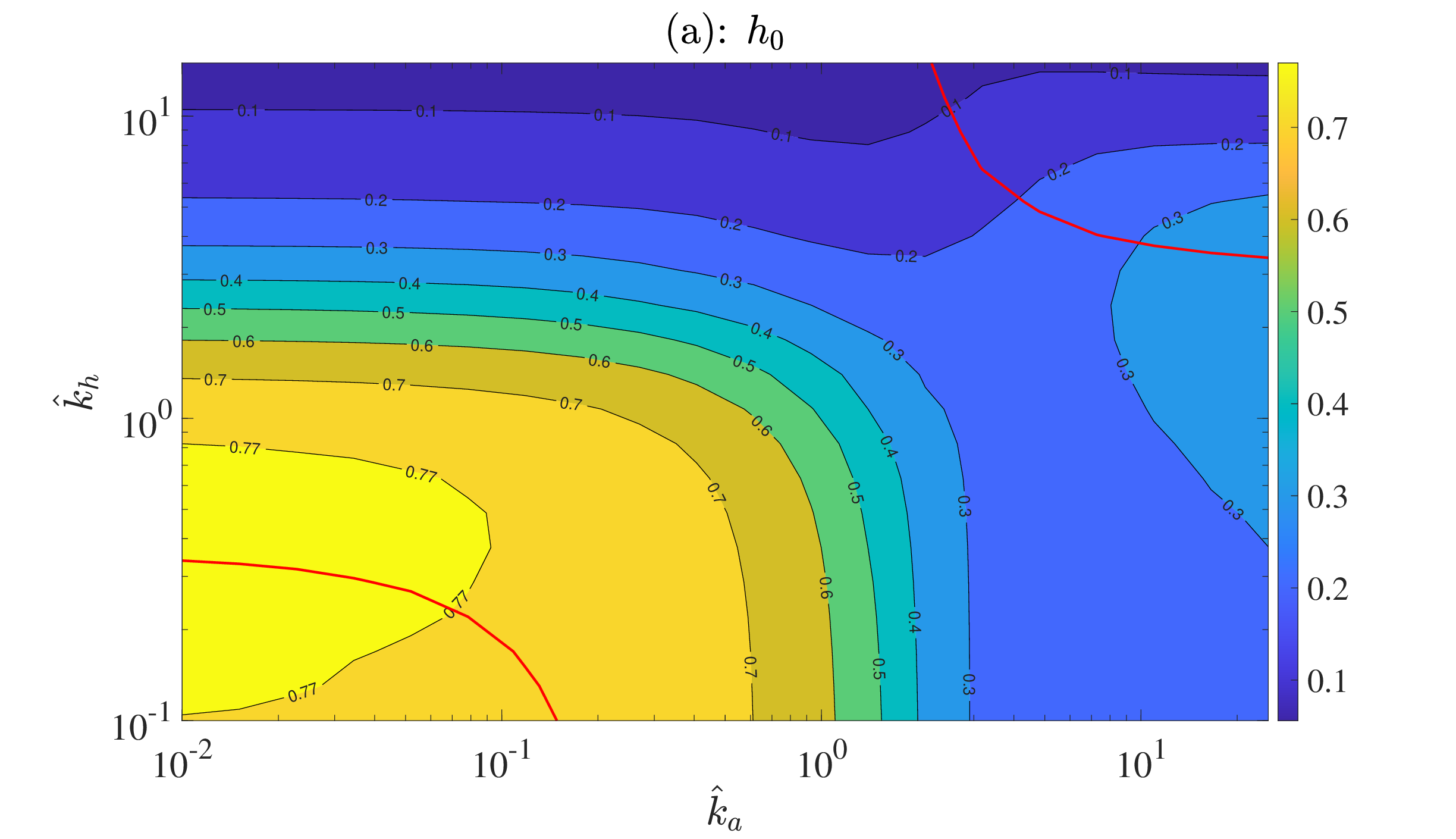,width=.55\linewidth}\epsfig{file=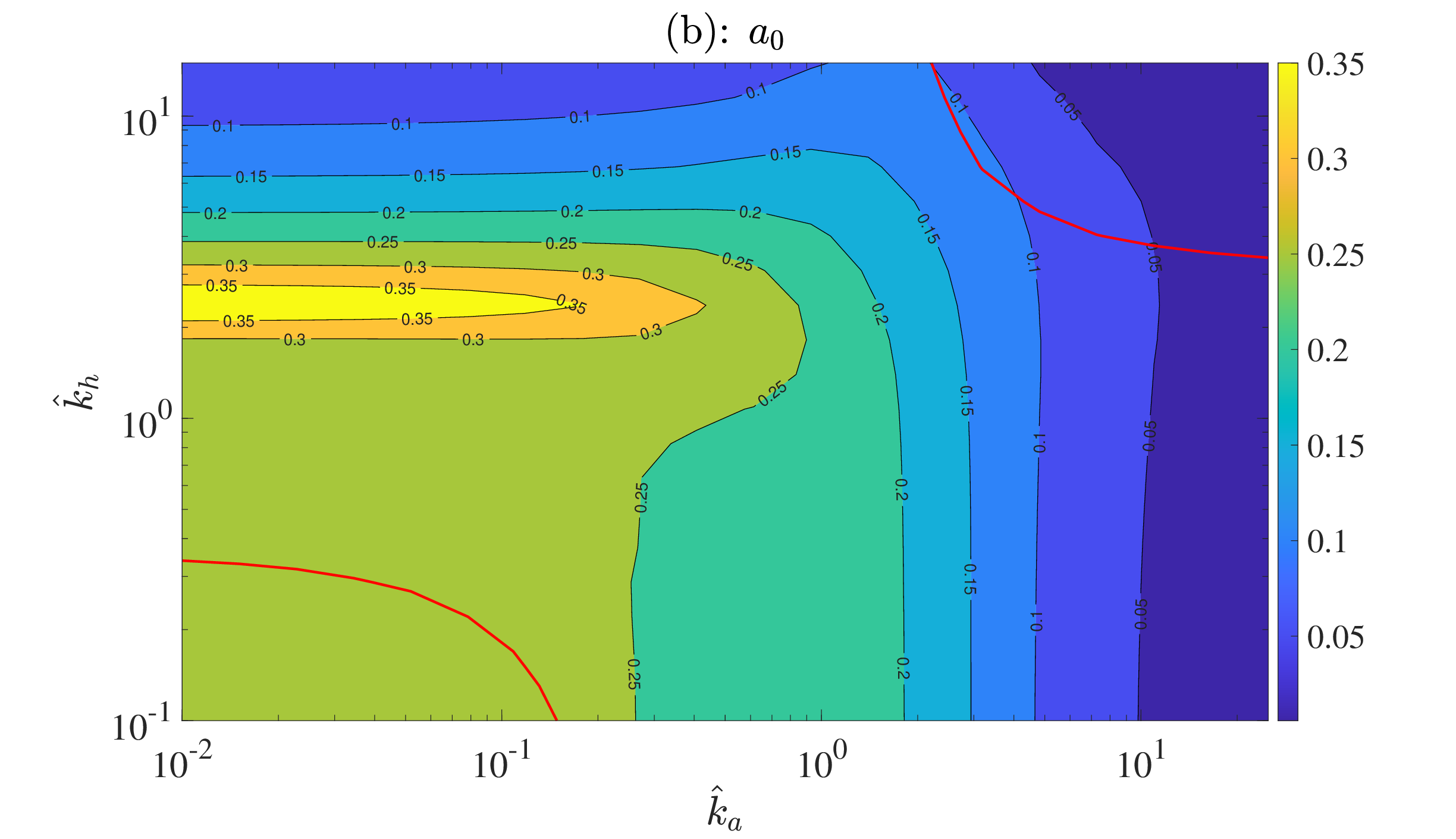,width=.55\linewidth}}
\centerline{\epsfig{file=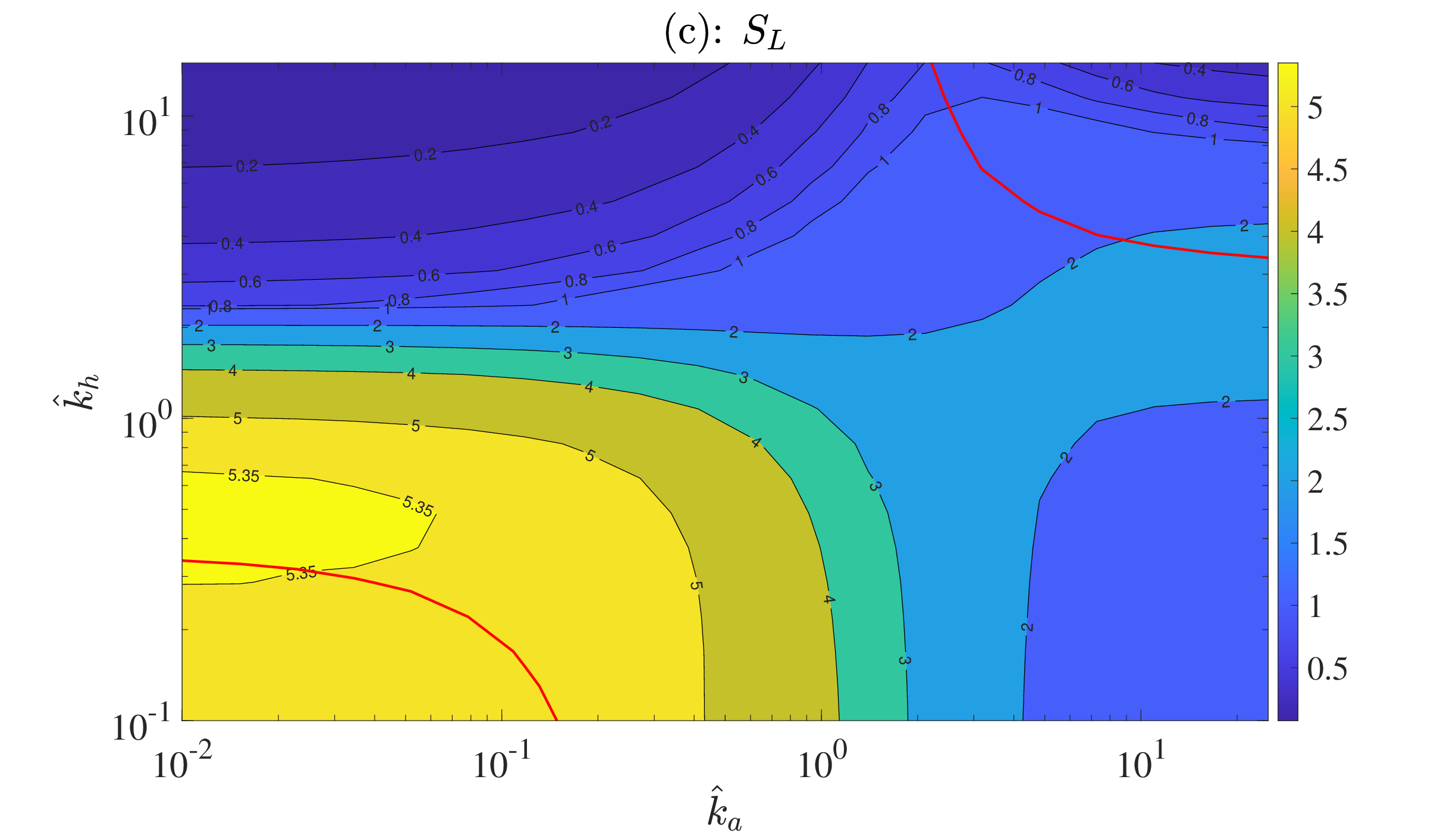,width=.55\linewidth}\epsfig{file=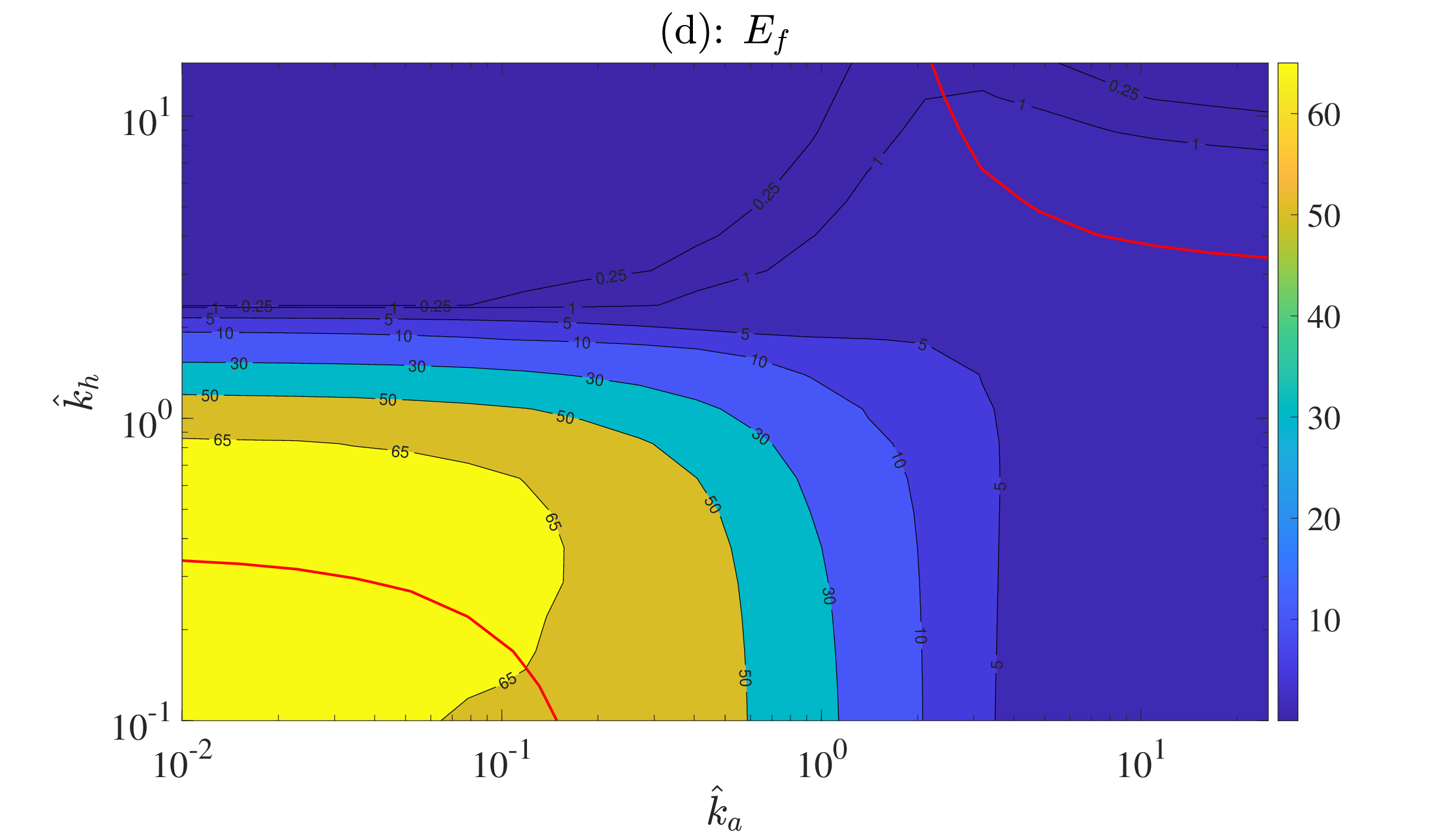,width=.55\linewidth}}
  \caption{As in Fig. \ref{fig_rigid3}, but for $R=0$.}
\label{fig_rigid4}
\end{figure}

Therefore, according to the present theory, the highest locomotion efficiency, both in terms of stride length and in terms of the efficiency \eqref{EFF},  is reached near the  springs resonance for $\hk_h$ slightly smaller than unity and for small $\hk_a$. Note that the pitch and heave amplitudes remain small for the selected value of the input force amplitude $\hA_l=1$, although close to the  validity range when $\hk_a$ is very small. A huge value of $E_f$ is obtained because  the pronounced maximum of $S_L$ approximately coincides with  a no less pronounced  minimum of the input power $\oC_P$ at these conditions. There is another local maximum of $S_L$ for $\hk_h \approx 2.4$ and $\hk_a$ of order of ten, associated to the torsional spring branch of the springs  resonance,  as  commented on above (corresponding to the one plotted in Fig. \ref{fig_rigid2} for $\hk_h=10$), but it is much smaller.

The locomotion performance pattern in the $(\hk_a,\hk_h)-$plane shown in Fig. \ref{fig_rigid3} remains in a similar fashion for other values of $R$, within the  range of small values of interest for aquatic propulsion  ($R \lesssim 1$). As $R$ decreases, the maxima of $h_0$, $a_0$, $S_L$ and $E_f$ for small values of $\hk_a$ and $\hk_k$ associated to the linear spring mode just become slightly smaller, as illustrated in Fig. \ref{fig_rigid4} for $R=0$ when compared with Fig. \ref{fig_rigid3}. However, the local maxima  for larger values of $\hk_a$ and $\hk_k$ associated to the torsional spring mode remains practically the same as $R$ decreases, just increasing very slightly.

Finally, in relation to the drag coefficient $\oC_D$, which is the only one remaining relevant parameter for a rigid foil, the propulsion patterns as $\oC_D$ varies  are also very similar to those shown in Figs. \ref{fig_rigid3} and \ref{fig_rigid4}. Obviously, as $\oC_D$ increases, the heave and pitch amplitudes have to increase to generate the higher  thrust needed to equate the drag, but the corresponding maxima of $S_L$ and $E_f$ decrease  with the increase of $\oC_D$ because the corresponding frequency and input power also increase. Above a certain value of $\oC_D$, the present linear theory may fail because the values of $h_0$ and $\alpha_0$ necessary to generate the corresponding thrust become too high. But, as seen in relation to the results of Fig. \ref{fig_zhong}c discussed above, even for fairly large values of $h_0$ the linear theory captures  quite well the experimental data.

\begin{figure}
\centerline{\epsfig{file=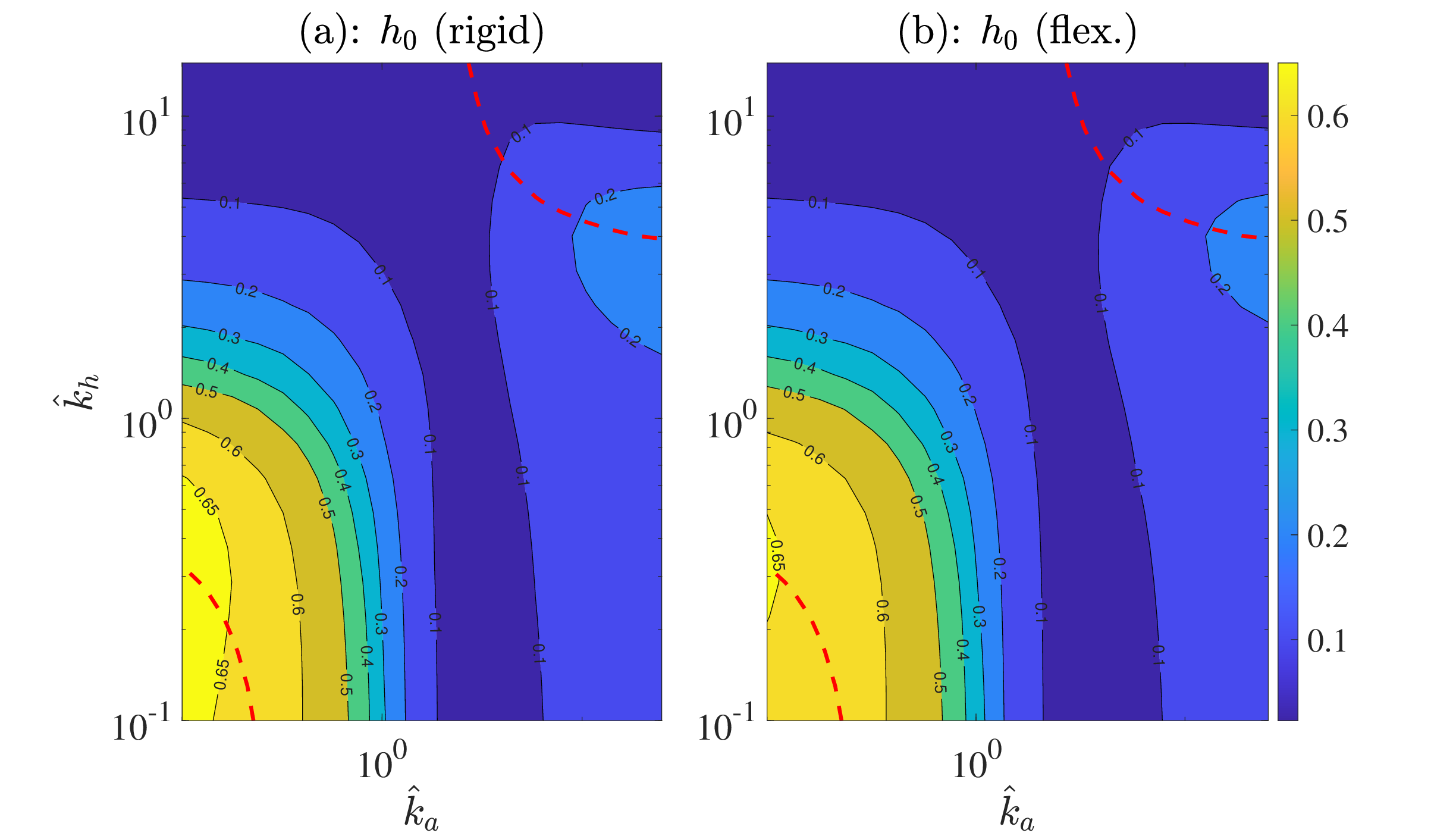,width=.55\linewidth}\epsfig{file=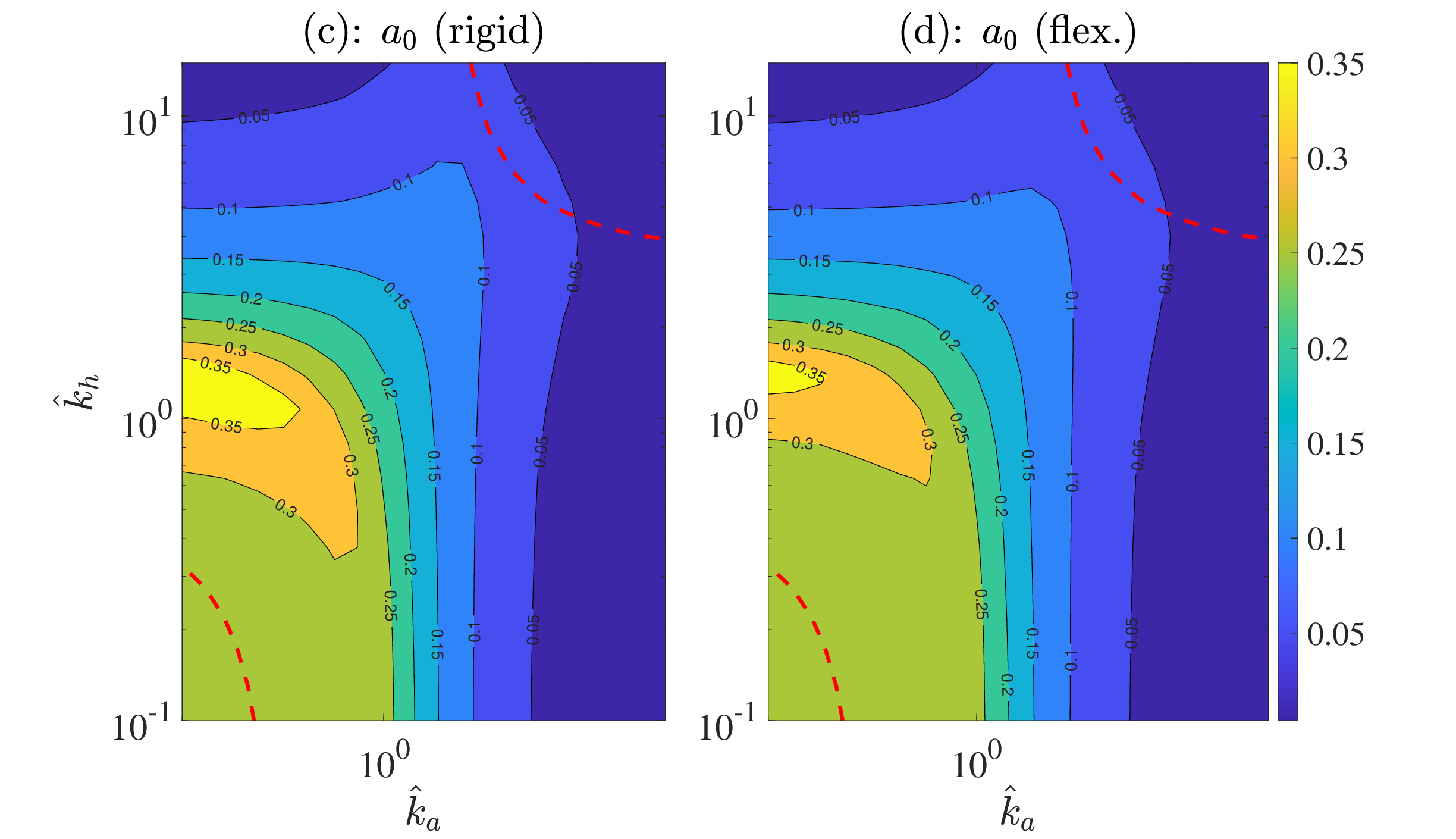,width=.55\linewidth}}
\centerline{\epsfig{file=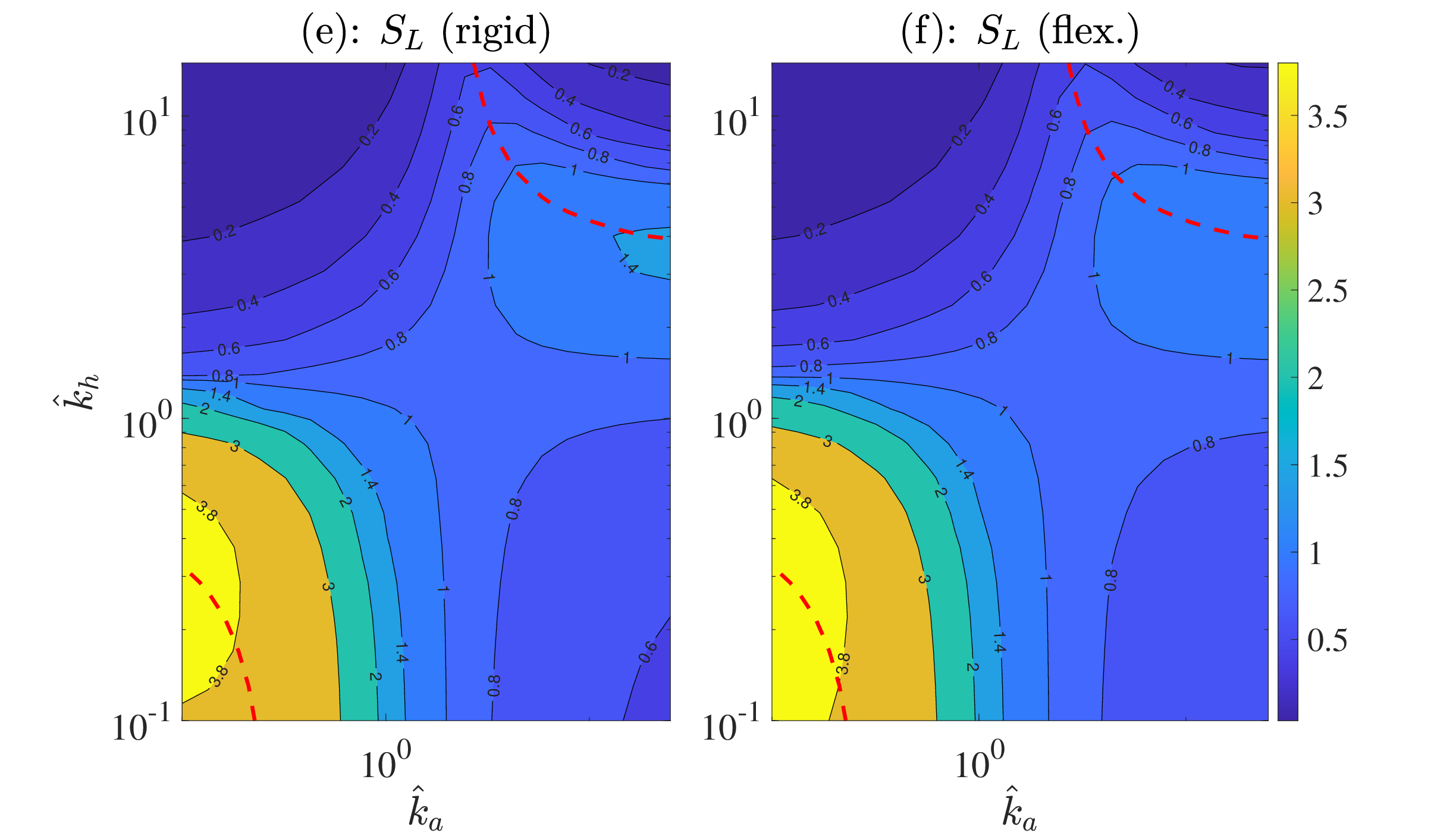,width=.55\linewidth}\epsfig{file=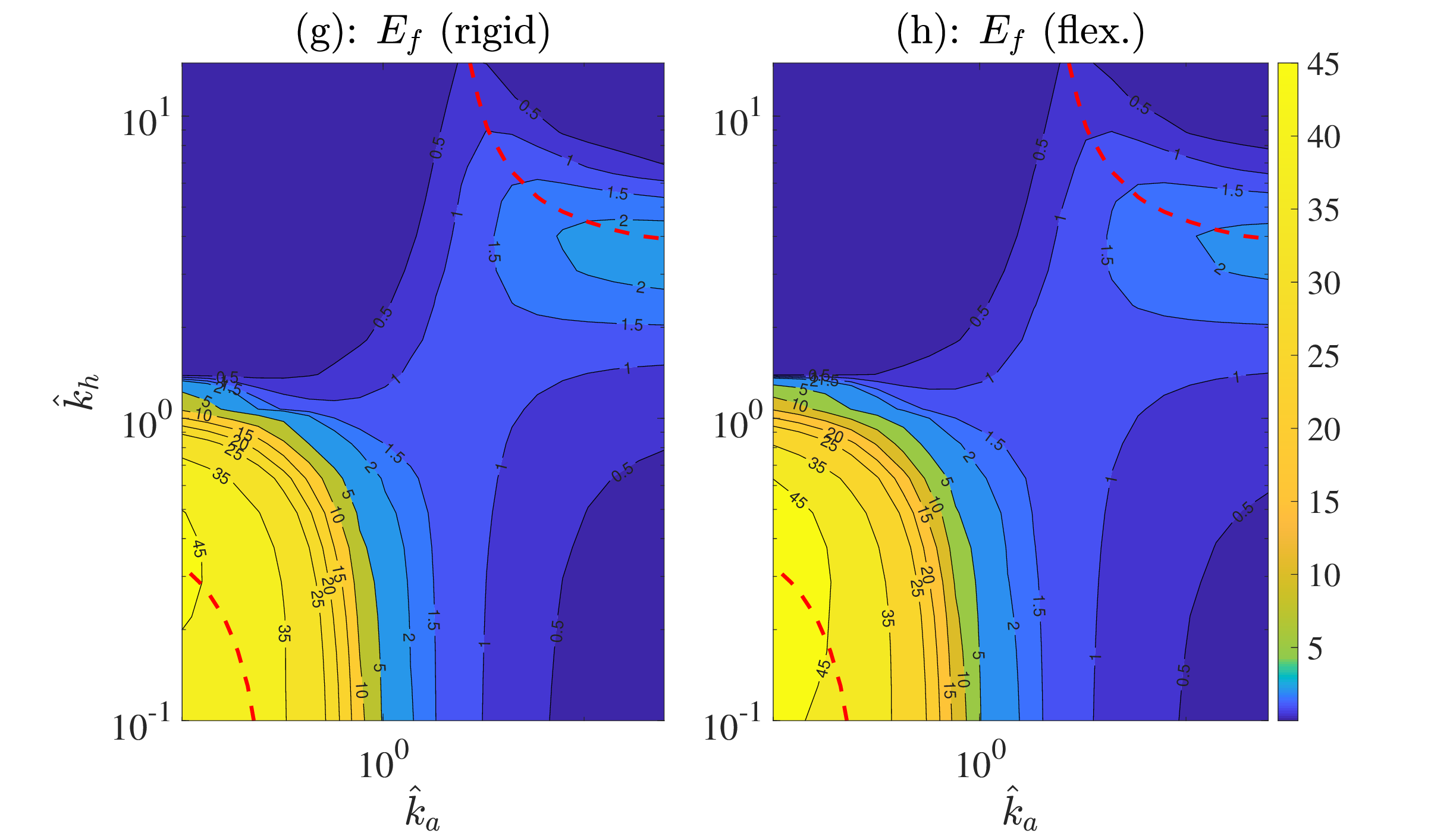,width=.55\linewidth}}
\centerline{\epsfig{file=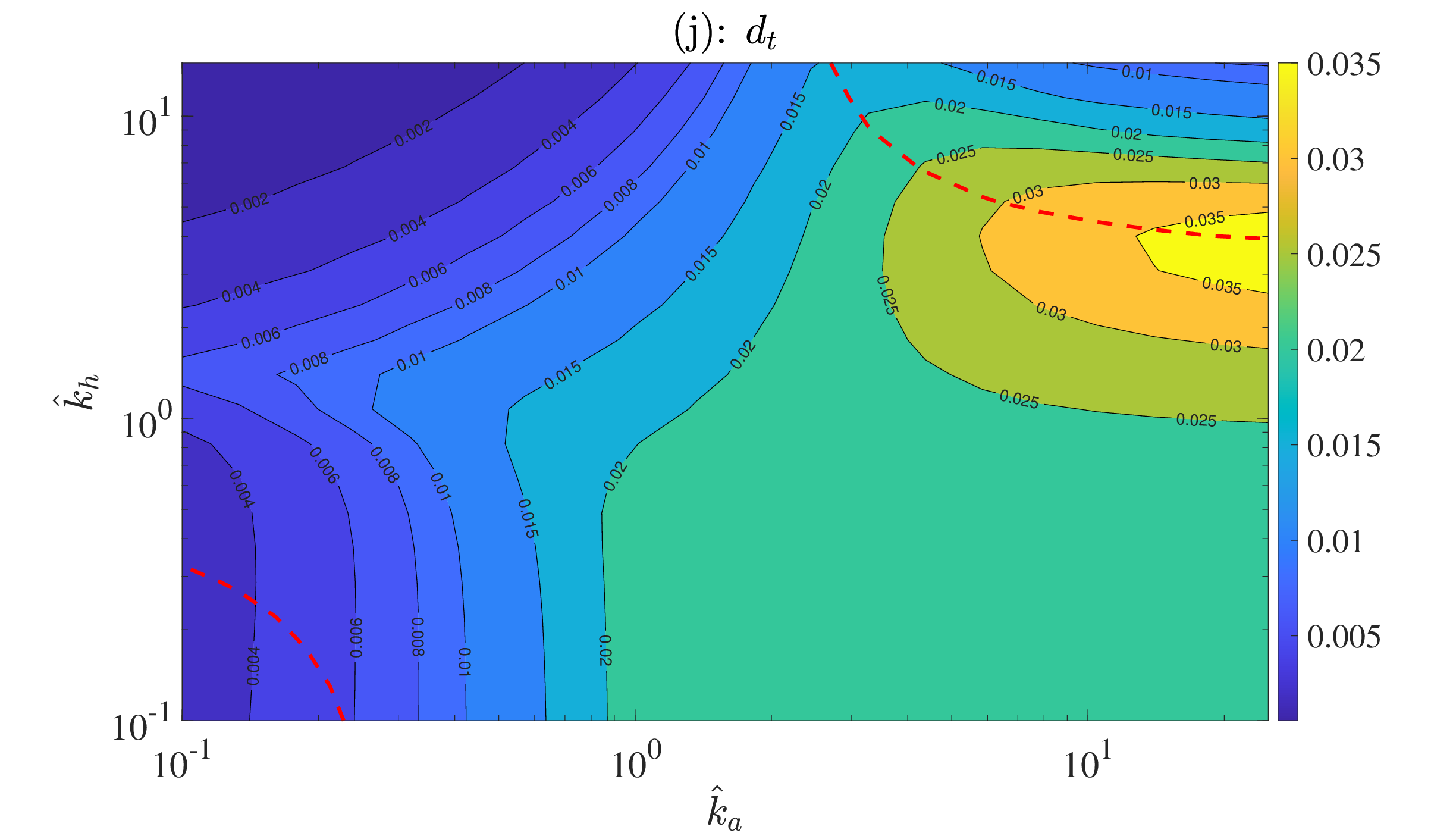,width=.55\linewidth}}
  \caption{Comparison in the $(\hk_a,\hk_h)-$plane of the heave (a-b) and pitch (c-d) amplitudes, stride length (e-f),  and efficiency (g-h)  for a rigid foil ($\hS=2\times 10^5$, (a,c,e,g)) and a flexible foil with $\hS=150$ (b,d,f,h), labelled with `rigid' and `flex.', respectively, on top of each subfigure.  $R=0.48$,  $\hb_a=\hb_h=0.05$,  $\hA_l =0.5$ and $\oC_D=0.25$. (j): Flexural defelection amplitude for the flexible foil. The red dashed lines correspond to the approximation of the resonant frequencies   \eqref{kr0ah} for a rigid foil.}
\label{fig_flex1}
\end{figure}

\section{Locomotion by a flexible foil with passive pitch and heave}
\label{sec_res_fle}

To see how the flexibility of the foil -- which in the present model is uniformly distributed along the chord length and it is independent of the local flexibility at the leading edge, characterized by a torsional spring that  allows for passive pitch -- affects to the locomotion performance, Fig. \ref{fig_flex1} compares in the $(\hk_a,\hk_h)-$plane some self-propulsion magnitudes for a rigid foil (the stiffness $\hS=2\times 10^5$ is used) and for a flexible foil with $\hS=150$, labelled as `rigid' and `flex.', respectively, on top of each subfigure, maintaining the same values  of the remaining non-dimensional parameters used in Figs. \ref{fig_rigid2} and \ref{fig_rigid3} (i.e., $R=0.48$,  $\hb_a=\hb_h=0.05$, $\oC_D=0.25$), except for $\hA_l$, which is selected  $0.5$ instead of unity.  
In addition to the comparison between the heave a pitch amplitudes, $h_0$ and $a_0$, the stride length $S_L$ and the efficiency $E_f$, 
Fig. \ref{fig_flex1} also shows the non-dimensional flexural deflection amplitude, i.e. (see Eq. \eqref{strouhal}),
\be d_t = |48 d_{10}+ 352 d_{20}| \,,  \label{DTT}
\ee
for the flexible foil.

The main feature to highlight in Fig. \ref{fig_flex1}  is that flexibility enhances the maxima of the stride length and the efficiency associated to the translational spring mode appearing for small $\hk_a$ and $\hk_h$, but reduces slightly the local maxima of  these magnitudes associated to the torsional spring mode for $\hk_a$ and $\hk_h$ of order of tens (see Figs. \ref{fig_flex1} e-h).  It also is noticeable that the approximation \eqref{kr0ah} for the springs natural frequencies of a rigid foil, marked with dashed red lines in Fig. \ref{fig_flex1}, still allows to estimate with enough precision the locations 
in which these local maxima occur in the $(\hk_a,\hk_h)-$plane even for a flexible foil. The local maxima of the heave and pitch amplitudes corresponding to both springs natural modes are somewhat smaller for the flexible foil than for the rigid foil  (Figs. \ref{fig_flex1} a-d). Thus, the deformation of the flexible  foil, whose flexural deflection amplitude is depicted in Fig. \ref{fig_flex1} j, allows to get a given  thrust ($\oC_T=\oC_D =0.25$ in the figure) with smaller pitch and heave amplitudes than for a rigid foil, and with the possibility of generating a more efficient locomotion that the rigid foil when actuated near the translational spring mode. This is perhaps the main achievement of the present configuration that permits passive heave when actuated by a given driving  force. 

That the best propulsion performance when only passive pitch is allowed is attained with a rigid foil elastically mounted through a torsional spring was already found in the pioneering work by \cite{moore15}. In the present configuration, that also includes passive heave and deformation, a better propulsion performance with more flexibility is attained associated to the translational branch of the springs natural modes. In fact,  Fig. \ref{fig_flex1} j shows that the maximum flexural deflection amplitude is associated to the torsional spring mode, meaning that in this mode  
deformation decreases propulsion performance. 

\begin{figure}
\centerline{\epsfig{file=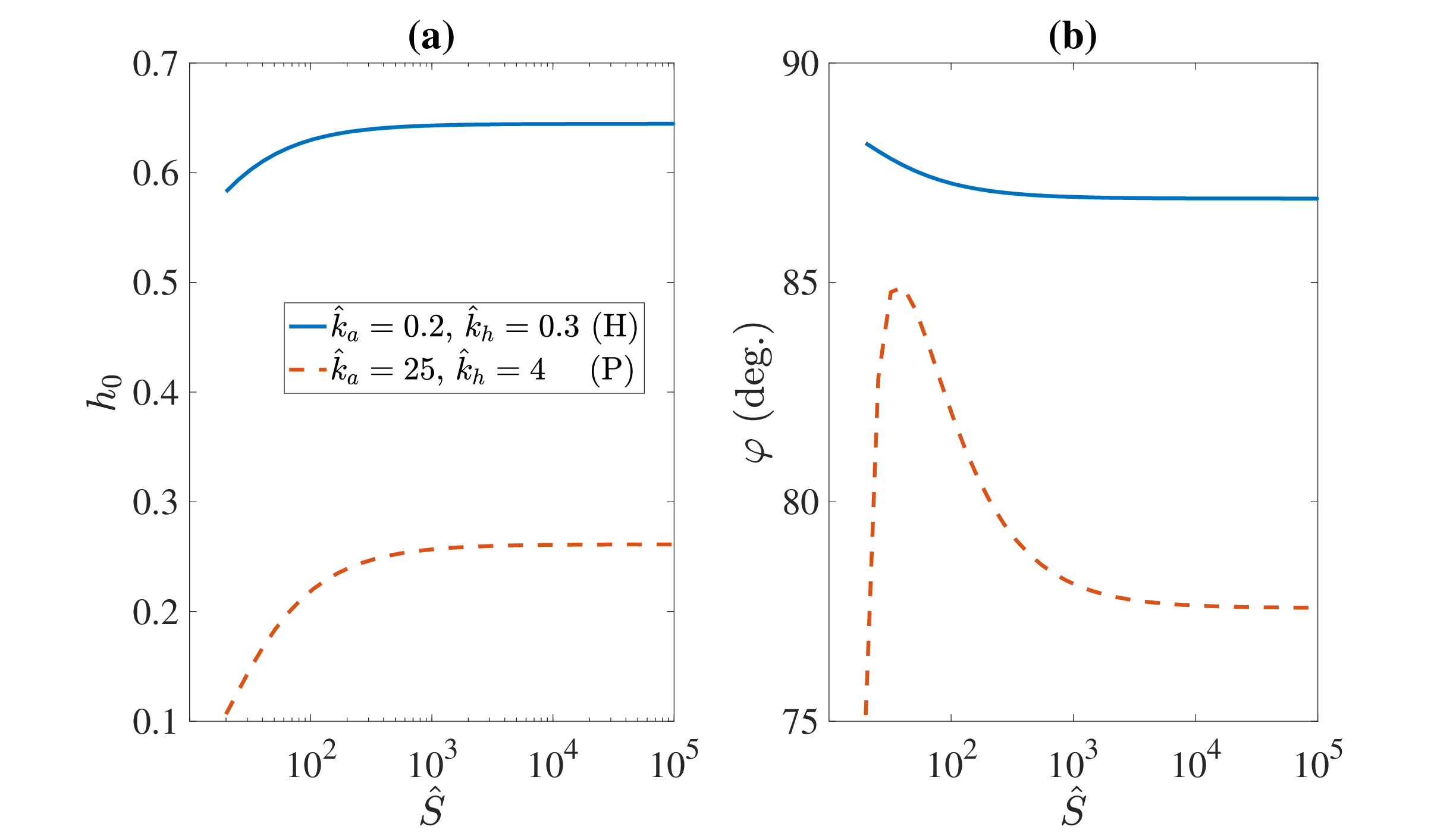,width=.55\linewidth}\epsfig{file=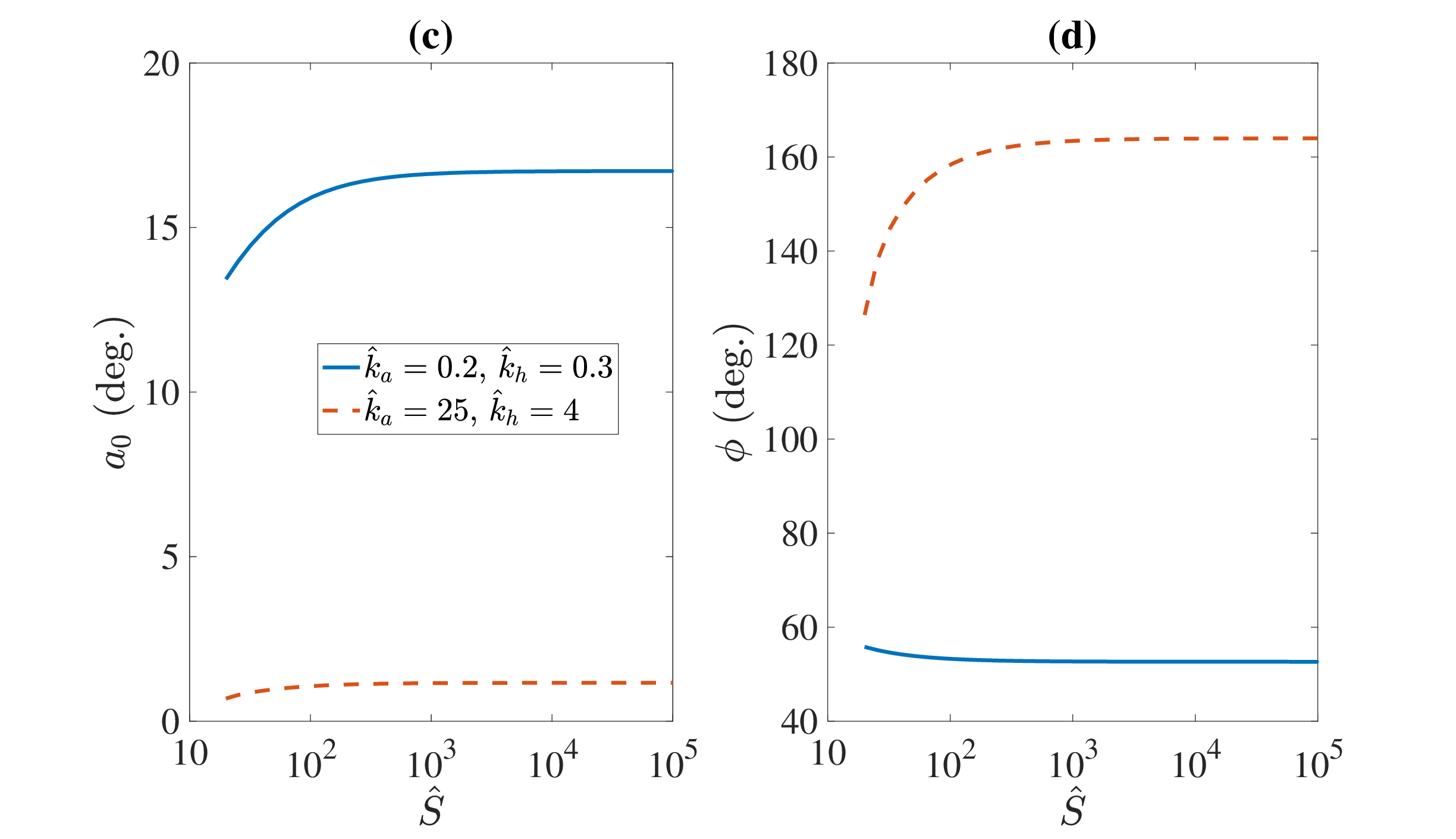,width=.55\linewidth}}
\centerline{\epsfig{file=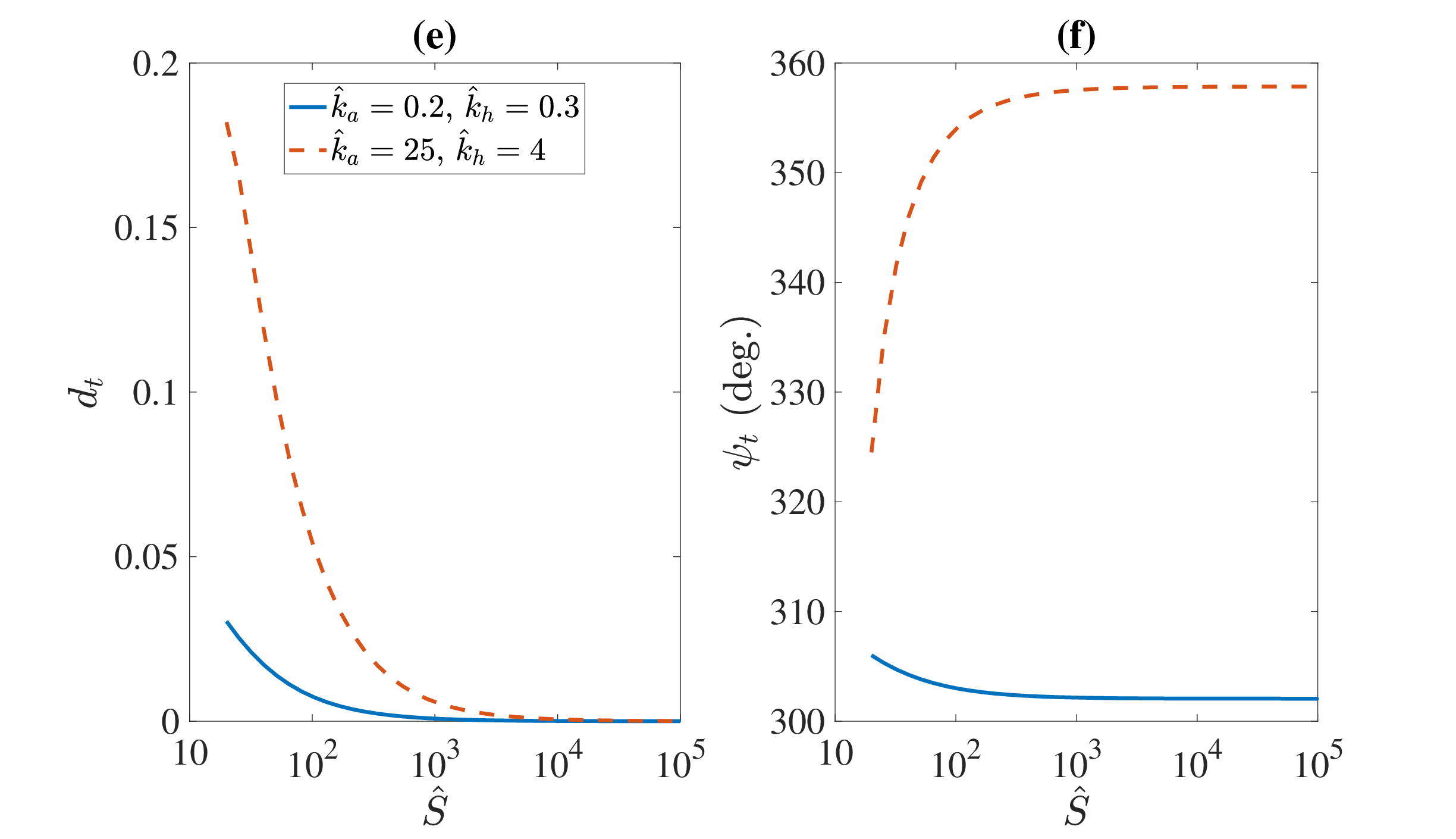,width=.55\linewidth}\epsfig{file=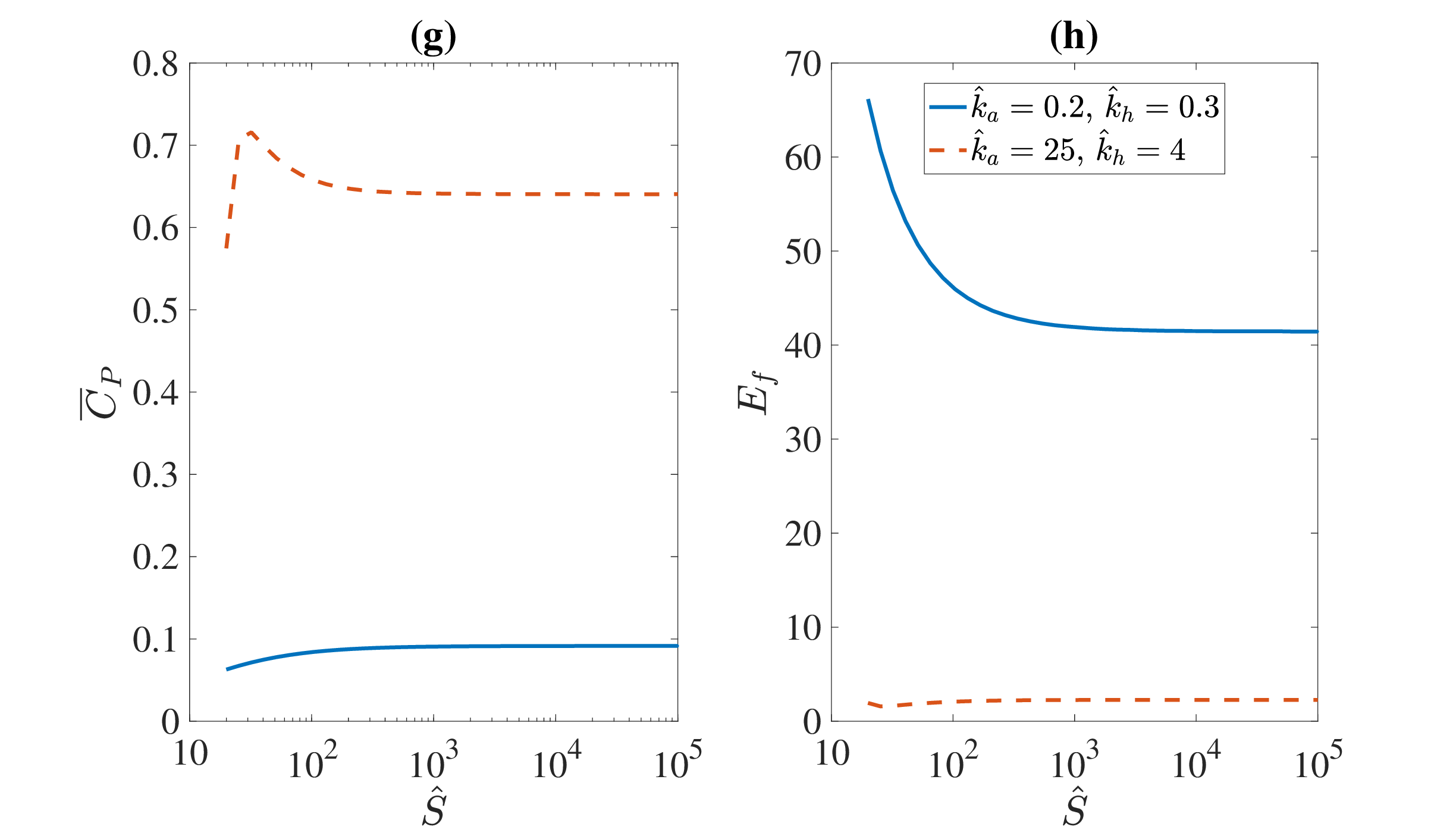,width=.55\linewidth}}
\centerline{\epsfig{file=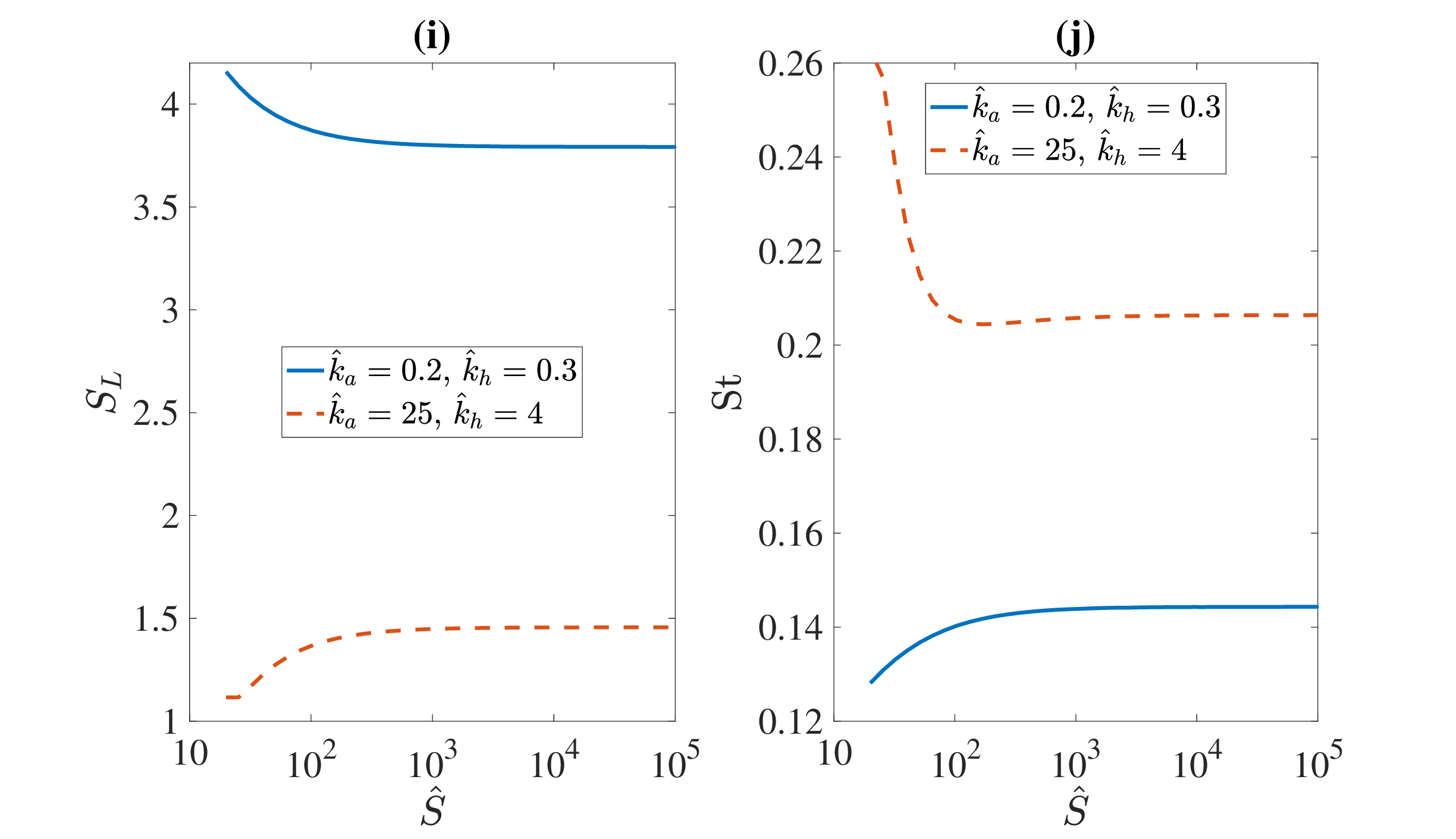,width=.55\linewidth}}
  \caption{Evolution with the rigidity $\hS$ of the heave amplitude (a) and phase (b), the pitch amplitude (c) and phase (d), the flexural deflection amplitude (e) and phase (f), the power input (g) and the efficiency (h), and the stride length (i) and  the Strouhal number (j) for the two pairs of optimal values of $\hk_a$ and $\hk_h$ represented in Fig. \ref{fig_flex1}, as indicated in some subfigures.   The remaining parameters are $R=0.48$,  $\hb_a=\hb_h=0.05$, $\hA_l =0.5$ and $\oC_D=0.25$.}
\label{fig_flex2}
\end{figure}

To better appreciate the trends  of optimal  locomotion performance near the two branches of the springs natural frequencies as the foil stiffness vary, Fig. \ref{fig_flex2} shows the evolution with  $\hS$ of the main non-dimensional parameters for $\hat{k}_a=0.2$ and $\hat{k}_h=0.3$, corresponding to the translational spring branch, and $\hat{k}_a=25$ with $\hat{k}_h=4$, corresponding to the torsional spring branch. To simplify the notation, these to modes will be dubbed, respectively, (H), associated to heave,  and (P), associated to pitch. Figures \ref{fig_flex2}h and \ref{fig_flex2}i show how the stride length and the efficiency increase as the stiffness $\hS$  of the foil decreases for the mode (H), while they increase with $\hS$, being both maxima for a rigid foil for the mode (P). But with the important difference that $S_L$ and $E_f$ are always much larger for the mode  (H) than for the mode (P). The heave and pitch amplitudes always increase with $\hS$, when, obviously, the flexural deflection amplitude decreases (Figs. \ref{fig_flex2} a, c and e). The foil deformation is significantly larger in the mode (P) than in the mode (H) as $\hS$ decreases, as commented on above. The phases of heave, pitch and deformation have a mixed behaviour in modes (H) and (P) as $\hS$ changes (Figs. \ref{fig_flex2} b, d and f). Finally, the Strouhal number (Fig. \ref{fig_flex2} j) is larger and has a larger variation with $\hS$ in the (P) mode than in the (H) mode.

\section{Concluding remarks}
\label{sec_conclu}

The bioinspired aquatic locomotion model developed here, based on the thrust generated by a flexible foil driven by a transversal oscillating force on  its leading edge, which is elastically attached to a virtual body responsible for the entire drag, so that the foil may undergo passive heaving and pitching oscillations in addition to passive deformation, includes as a limiting case  previous similar models  for aquatic bioinspired self-propulsion using a rigid foil, reproducing their  results. Particularly, the model includes that developed by \cite{zhozh21} for a rigid foil  actuated by a given driving heaving motion and attached to the virtual body through  a torsional spring with variable stiffness simulating the variable muscle tension of the peduncle joining the caudal fin to the trunk of a tuna-like fish or robot. The results obtained by these authors for the the stride length with this simple model, which are reproduced by the present model,  capture and  explain remarkably well the main features of  their experimental results with tuna-inspired robots, thus validating the present model in this limit.

The inclusion of passive heave and deformation of the foil when actuated by a given driving force on its leading edge 
increases the possibility of a better locomotion performance by enlarging the parameters space. Thus, still for a rigid foil elastically mounted, in addition to the optimal propulsion performance associated to the torsional spring resonance of the previous models, now there exists a new optimal parametric region of even better propulsion performance (with larger stride lengths and locomotion efficiencies) around the translational spring branch of the coupled  translational and torsional springs resonance. These two parametric  regions in the springs constants $(\hk_a,\hk_h)$ plane where the propulsion performance reaches these two local maxima are well characterized by an analytical approximation of the resonant frequency of the coupled spring modes given by Eq.  \eqref{kr0ah}, with takes into account the effects of inertia of the foil and the main contributions of the hydrodynamic force and moment. One local maximum is around the torsional spring branch for large values of $\hk_a$, and the other one, with higher stride lengths and propulsion efficiencies, around the translational spring branch for small $\hk_a$. 

When the flexibility of the foil is taken into account, the optimal propulsion performance of the translational spring branch is enhanced as the stiffness of the foil decreases (down to the validity limit of the present model), whereas  the optimal propulsion performance associated to the torsional spring branch slowly grows as the stiffness increases, reaching a maximum stride length for a rigid foil, as already known from previous models. The best self-propulsion performance found with the present model is thus attained by a not too rigid foil ($\hS$ below $10^2$) for small values of $\hk_a$ and $\hk_h$ located around the translational spring branch of the springs resonant frequency. It is remarkable that the approximation  \eqref{kr0ah} for the springs resonant frequencies of a rigid foil remains a good approximation as the stiffness decreases, within the limits of the present model. In any case, the present analytical model is a useful tool to characterize the best self-propulsion performance for any set of given parameters, provided that the resulting pitch, heave and deformation amplitudes remain small.\\

\backsection[Funding]{
This research has been supported by the MCIyU/AEI grant PID2023-150588NB-I00 and by the Junta de Andaluc\'ia grant PPRO-TEP146-G-FEDER-2023. }%

\backsection[Declaration of interests]{The authors report no conflict of interest.}

\appendix

\numberwithin{equation}{section}
\section{Fluid force and moment coefficients}
\label{app_coeffi}
In the present linear potential flow approximation the fluid force and moment coefficients are given, for the foil motion \eqref{zs_def} and harmonic functions $h(t)$, $\alpha(t)$, $d_1(t)$ and $d_2(t)$ with reduced frequency
\be k = \frac{\omega  c}{2 U} \,, 
\ee
by \citep{ferna25}
\be C_{L}(t) =   \pi \left( \dot{\alpha} + \ddot{\alpha} - \ddot{h} - 25 \dt_1 - \frac{149}{8} \ddt_1 - \frac{1465}{8} \dt_2 - \frac{1073}{8} \ddt_2 \right) +  \Gamma_0(t) \C (k) \,, \label{CL0}
\ee
\[ C_{M}(t)= \frac{\pi}{256} \left( 192 \dot{\alpha} + 144 \ddot{\alpha} - 128 \ddot{h} - 576 d_1 -5248 \dt_1 - 2800 \ddt_1 - 4640 d_2  \right. \]
\be \left. - 38680 \dt_2-20213 \ddt_2 \right) + \frac{1}{4} \Gamma_0 (t) \C(k) \,, \label{CM0}
\ee
\[ C_{F_{1}}(t)= \frac{\pi}{192} \left( 384 \dot{\alpha} + 288 \ddot{\alpha} - 240 \ddot{h} - 1056 d_1 -10848 \dt_1 - 5745  \ddt_1 - 8640 d_2 \right. \]
\be \left.  - 80190 \dt_2-41540 \ddt_2 \right) +  \frac{1}{2} \Gamma_0 (t) \C(k) \,, \label{CF10}
\ee
\[ C_{F_{2}}(t)= \frac{\pi}{128} \left( 368 \dot{\alpha} + 280 \ddot{\alpha} - 224 \ddot{h} - 912 d_1 -10580 \dt_1 - 5680 \ddt_1 - 7530 d_2 \right. \]
\be \left.  - 78350 \dt_2-41117 \ddt_2 \right) +  \frac{5}{8} \Gamma_0 (t) \C(k)  \,, \label{CF20}
\ee
with
\be \Gamma_0(t)= \pi \left[ - 2 \dot{h} + 3 \dot{\alpha} + 2 \alpha - \frac{263}{4} \dt_1 - 59 d_1 - \frac{3831}{8} \dt_2 - \frac{1755}{4} d_2 \right] \,, \label{G0t}
\ee
and $\C(k)$ is Theodorsen's function. Note that now the reduced frequency $k$  is an unknown of the problem. 

Terms containing $\C(k)$ correspond to the circulatory part of the force and moments, while the remaining ones  correspond to the non-circulatory (or added-mass) contributions.
For a rigid foil, $d_{1}=d_2=0$ and Eqs. \eqref{CL0}  and \eqref{CM0} with \eqref{G0t} coincide with the classical expressions derived   by \cite{theod35} for the lift and moment of a  heaving and pitching foil about its leading edge \citep[see also][]{garri36}.

\section{Terms in matrix $\tA$ in Eq,  \eqref{def_eq_Ad0} and $\hat{\tA}$ in Eq. \eqref{sistema}}
\label{app_coeffA}

Substituting the force and moment expression given in Appendix \ref{app_coeffi} for the harmonic motion \eqref{harmo}  into Eqs. \eqref{mom1a}, \eqref{mom2a}, \eqref{mom3a} and \eqref{mom4a}, the matrices in the resulting system of linear equations \eqref{def_eq_Ad0} are the following:

\be \tA_0 = \left( \begin{array}{cccc}  -k^2 R+k_h+ i k b_h   &  k^2 R   &     -96 k^2 R/5   &    -416 k^2 R/3 \\
    -k^2 R/2      &        2  k^2 R/3-k_a- i k b_a  & -208  k^2 R/15   &  -704  k^2 R/7 \\
    -4  k^2 R/3     &       2  k^2 R         &       32  S/3- 4544  k^2 R/105   &  80  S-944  k^2 R/3 \\
    -2  k^2 R        &      16  k^2 R/5       &      16  S- 496  k^2 R/7   &  128  S- 32512  k^2 R/63
\end{array} \right) \,, \label{def_A0}
\ee

\be \tA_f =  \pi \left( \begin{array}{cccc}  l_h   &  l_a   &     l_1   &    l_2 \\
    m_h     &        m_a  & m_1  & m_2 \\
  f_{1h}    &       f_{1a}        &      f_{11}   &  f_{22} \\
   f_{2h}        &    f_{2a}     &     f_{21}   &  f_{22}
\end{array} \right) \,,  \label{Af}
\ee
with
\be l_h =  -k^2 +2 i k \C(k)  \,, \label{ahh} 
\ee
\be l_a = -  ik +k^2 -\C(k) \left(3 ik + 2 \right)   \,, 
\ee
\be l_1 =  25 i k  - \frac{149}{8} k^2   + \C(k) \left( \frac{263}{4}  i k + 59 \right)   \,, 
\ee
\be l_2 =  \frac{1465}{8} i k  - \frac{1073}{8} k^2   + \C(k) \left( \frac{3831}{8}  i k + \frac{1755}{4} \right)  \,, 
\ee
\be m_h =  - \frac{1}{2} k^2 + \frac{1}{2} i k \C(k)   \,,
\ee
\be m_a = - \frac{3}{4} ik + \frac{9}{16} k^2 - \frac{1}{4} \C(k) \left(3 ik + 2 \right)  \,, 
\ee
\be m_1 =  \frac{9}{4} + \frac{1321}{64} i k  - \frac{175}{16} k^2   + \frac{1}{4} \C(k) \left( \frac{263}{4}  i k + 59 \right)    \,, 
\ee
\be m_2 =  \frac{145}{8} +\frac{4835}{32} i k  - \frac{20213}{256} k^2   + \frac{1}{4} \C(k) \left( \frac{3831}{8}  i k + \frac{1755}{4} \right)  \,, 
\ee
\be f_{1h} =  - \frac{5}{4} k^2 + i k \C(k)   \,,
\ee
\be f_{1a} =  - 2 ik + \frac{3}{2} k^2 - \frac{1}{2} \C(k) \left(3 ik + 2 \right)   \,, 
\ee
\be f_{11} =  \frac{11}{2} + \frac{113}{2} i k  - \frac{5745}{192} k^2   + \frac{1}{2} \C(k) \left( \frac{263}{4}  i k + 59 \right)    \,, 
\ee
\be f_{12} = 45 +\frac{13365}{32} i k  - \frac{10385}{48} k^2   + \frac{1}{2} \C(k) \left( \frac{3831}{8}  i k + \frac{1755}{4} \right)  \,, 
\ee
\be f_{2h} =  - \frac{7}{4} k^2 + \frac{5}{4}  i k \C(k)   \,,
\ee
\be f_{2a} =  - \frac{23}{8} ik + \frac{35}{16} k^2 - \frac{5}{8} \C(k) \left(3 ik + 2 \right)   \,, 
\ee
\be f_{21} =  \frac{57}{8} + \frac{2645}{32} i k  - \frac{355}{8} k^2   + \frac{5}{8} \C(k) \left( \frac{263}{4}  i k + 59 \right)    \,, 
\ee
\be f_{22} =  \frac{3765}{64} +\frac{39175}{64} i k  - \frac{41117}{128} k^2   + \frac{5}{8} \C(k) \left( \frac{3831}{8}  i k + \frac{1755}{4} \right)  \,. \label{a22} 
\ee

On the other hand, the `hatted' matrices in \eqref{sistema} are the above ones divided by $k^2$ with $S$, $k_h$,  $b_h$, $k_a$ and $b_a$ replaced by their hatted counterparts. Particularly, $\hat{\tA}_0$ does not depend on $k$,
\be \hat{\tA}_0 = \left( \begin{array}{cccc}  - R+\hk_h+ i \hb_h   &   R   &     -96  R/5   &    -416  R/3 \\
    - R/2      &        2   R/3-\hk_a- i \hb_a  & -208   R/15   &  -704   R/7 \\
    -4  R/3     &       2   R         &       32  \hS/3- 4544   R/105   &  80  \hS-944   R/3 \\
    -2   R        &      16   R/5       &      16  \hS- 496   R/7   &  128  \hS- 32512   R/63
\end{array} \right) \,, \label{def_A0hat}
\ee
and $\hat{\tA}_f$ is straightforwardly obtained dividing $\tA_f$ by $k^2$.

\bibliographystyle{jfm}

\begin{thebibliography}{39}
\expandafter\ifx\csname natexlab\endcsname\relax\def\natexlab#1{#1}\fi
\def\au#1{#1} \def\ed#1{#1} \def\yr#1{#1}\def\at#1{#1}\def\jt#1{\textit{#1}}
  \def\bt#1{#1}\def\bvol#1{\textbf{#1}} \def\vol#1{#1} \def\pg#1{#1}
  \def\publ#1{#1}\def\arxiv#1{#1}\def\org#1{#1}\def\st#1{\textit{#1}}

\bibitem[Akoz \& Moored(2018)]{akomo18}
{\sc \au{Akoz, E.} \& \au{Moored, K.~W.}} \yr{2018}  \at{Unsteady propulsion by
  an intermittent swimming gait}.  \jt{J. Fluid Mech.}  \bvol{834},
  \pg{149--172}.

\bibitem[Alaminos-Quesada(2021)]{alami21}
{\sc \au{Alaminos-Quesada, J.}} \yr{2021}  \at{Limit of the two-dimensional
  linear potential theories on the propulsion of a flapping airfoil in forward
  flight in terms of the {Reynolds and Strouhal} number}.  \jt{Phys. Rev.
  Fluids}  \bvol{6}~(12),  \pg{123101}.

\bibitem[Anderson {\em et~al.\/}(1998)Anderson, Streitlien, Barret \&
  Triantafyllou]{andst98}
{\sc \au{Anderson, J.~M.}, \au{Streitlien, K.}, \au{Barret, K.~S.} \&
  \au{Triantafyllou, M.~S}} \yr{1998}  \at{Oscillating foils of high propulsive
  efficiency}.  \jt{J. Fluid Mech.}  \bvol{360},  \pg{41--72}.

\bibitem[Bockmann \& Steen(2014)]{bocst14}
{\sc \au{Bockmann, E.} \& \au{Steen, S.}} \yr{2014}  \at{Experiments with
  actively pitch-controlled and spring-loaded oscillating foils}.  \jt{Appl.
  Ocean Res.}  \bvol{48},  \pg{227--235}.

\bibitem[Eloy(2012)]{eloy12}
{\sc \au{Eloy, C.}} \yr{2012}  \at{Optimal {Strouhal} number for swimming
  animals}.  \jt{J. Fluids Structures}  \bvol{30},  \pg{205--218}.

\bibitem[Fernandez-Feria(2017)]{ferna17}
{\sc \au{Fernandez-Feria, R.}} \yr{2017}  \at{Note on optimum propulsion of
  heaving and pitching airfoils from linear potential theory}.  \jt{J. Fluid
  Mech.}  \bvol{826},  \pg{781--796}.

\bibitem[Fernandez-Feria(2025)]{ferna25}
{\sc \au{Fernandez-Feria, R.}} \yr{2025}  \at{Comparing analytically propulsion
  by pitching and heaving flexible foils near the first two natural modes}.
  \jt{J. Fluid Mech.}  \bvol{1015},  \pg{A35}.

\bibitem[Fernandez-Feria \& Alaminos-Quesada(2021)]{feral21a}
{\sc \au{Fernandez-Feria, R.} \& \au{Alaminos-Quesada, J.}} \yr{2021}
  \at{Propulsion and energy harvesting performances of a flexible thin airfoil
  undergoing forced heaving motion with passive pitching and deformation of
  small amplitude}.  \jt{J. Fluids Structures}  \bvol{102},  \pg{103255}.

\bibitem[Fish \& Lauder(2006)]{fisla06}
{\sc \au{Fish, F.~E.} \& \au{Lauder, G.~V.}} \yr{2006}  \at{Passive and active
  flow control by swimming fishes and mammals}.  \jt{Annu. Rev. Fluid. Mech.}
  \bvol{38},  \pg{193--224}.

\bibitem[Floryan \& Rowley(2018)]{floro18}
{\sc \au{Floryan, D.} \& \au{Rowley, C.~W.}} \yr{2018}  \at{Clarifying the
  relationship between efficiency and resonance for flexible inertial
  swimmers}.  \jt{J. Fluid Mech.}  \bvol{853},  \pg{271--300}.

\bibitem[Garrick(1936)]{garri36}
{\sc \au{Garrick, I.~E.}} \yr{1936}  \bt{Propulsion of a flapping and
  oscillating airfoil}. {\em Tech. Rep.\/} TR 567.  \org{NACA}.
  
 \bibitem[Graziani {\em et~al.\/}(2025)Graziani, Paniccia \& Piva]{grapa25}
{\sc \au{Graziani, G.}, \au{Paniccia, D.} \& \au{Piva, R.}} \yr{2025}
  \at{Aquatic locomotion due to a flexible foil flapping in a perfect fluid}.
  \jt{J. Fluid Mech.}  \bvol{1021},  \pg{A34}. 

\bibitem[Graziani {\em et~al.\/}(2026)Graziani, Paniccia \& Piva]{grapa26}
{\sc \au{Graziani, G.}, \au{Paniccia, D.} \& \au{Piva, R.}} \yr{2026}
  \at{Aquatic locomotion by a flapping tail with passive pitch and tunable
  stiffness}.  \jt{J. Fluids Structures}  \bvol{142},  \pg{104524}.

\bibitem[Hu {\em et~al.\/}(2025)Hu, Yang, Zhao, Wang, Liao \& Yu]{huyan25}
{\sc \au{Hu, H.}, \au{Yang, J.}, \au{Zhao, L.}, \au{Wang, Y.}, \au{Liao, W.} \&
  \au{Yu, W.}} \yr{2025}  \at{The relationship among resonance, thrust, and
  hydrodynamic efficiency of a torsional-spring-based passive pitching
  propulsor}.  \jt{Phys. Fluids}  \bvol{37}~(3),  \pg{031906}.

\bibitem[Lauder(2000)]{laude00}
{\sc \au{Lauder, G.~V.}} \yr{2000}  \at{Function of the caudal fin during
  locomotion in fishes: {Kinematics}, flow visualization, and evolutionary
  patterns}.  \jt{Am. Zool.}  \bvol{40}~(1),  \pg{101--122}.

\bibitem[Lauder(2015)]{laude15}
{\sc \au{Lauder, G.~V.}} \yr{2015}  \at{Fish locomotion: {Recent} advances and
  new directions}.  \jt{Annu. Rev. Mar. Sci.}  \bvol{7},  \pg{521--545}.

\bibitem[Lighthill(1970)]{light70}
{\sc \au{Lighthill, M.~J.}} \yr{1970}  \at{Aquatic animal propulsion of high
  hydromechanical efficiency}.  \jt{J. Fluid Mech.}  \bvol{44},  \pg{265--301}.

\bibitem[Lopez-Tello \& Fernandez-Feria(2023)]{lopfe23a}
{\sc \au{Lopez-Tello, P.~E.} \& \au{Fernandez-Feria}} \yr{2023}  \at{Effect of
  flexibility on the self-propelled locomotion by an elastically supported
  stiff foil actuated by a torque}.  \jt{Phys. Rev. Fluids}  \bvol{8},
  \pg{063102}.

\bibitem[Lucas {\em et~al.\/}(2014)Lucas, Johnson, Beaulieu, Cathcart, Tirrell,
  Colin, Gemmell, Dabiri \& Costello]{lucjo14}
{\sc \au{Lucas, "K.~N.}, \au{Johnson, N.}, \au{Beaulieu, W.~T.}, \au{Cathcart,
  E.}, \au{Tirrell, G.}, \au{Colin, S.~P.}, \au{Gemmell, B.~J.}, \au{Dabiri,
  J.~O.} \& \au{Costello, J.~H.}} \yr{2014}  \at{Bending rules for animal
  propulsion}.  \jt{Nat. Commun.}  \bvol{5},  \pg{3293}.

\bibitem[Mei {\em et~al.\/}(2018)Mei, Yan, Zhou, Guo, Cong \& Shi]{meiya23}
{\sc \au{Mei, L.}, \au{Yan, W.}, \au{Zhou, J.}, \au{Guo, B.}, \au{Cong, L.} \&
  \au{Shi, W.}} \yr{2018}  \at{Propulsion characteristics of self-pitching
  flapping foil}.  \jt{Ocean Eng.}  \bvol{285},  \pg{115233}.

\bibitem[Moore(2014)]{moore14}
{\sc \au{Moore, M. N.~J.}} \yr{2014}  \at{Analytical results on the role of
  flexibility in flapping propulsion}.  \jt{J. Fluid Mech.}  \bvol{757},
  \pg{599--612}.

\bibitem[Moore(2015)]{moore15}
{\sc \au{Moore, M. N.~J.}} \yr{2015}  \at{Torsional spring is the optimal
  flexibility arrangement for thrust production of a flapping wing}.  \jt{Phys.
  Fluids}  \bvol{27},  \pg{091701}.

\bibitem[Moored \& Quinn(2019)]{mooqu19}
{\sc \au{Moored, K.~W.} \& \au{Quinn, D.~B.}} \yr{2019}  \at{Inviscid scaling
  laws of a self-propelled pitching airfoil}.  \jt{AIAA J.}  \bvol{57}~(9),
  \pg{3686--3700}.

\bibitem[Paniccia {\em et~al.\/}(2021)Paniccia, Padovani, Graziani \&
  Piva]{panpa21}
{\sc \au{Paniccia, D.}, \au{Padovani, L.}, \au{Graziani, G.} \& \au{Piva, R.}}
  \yr{2021}  \at{The performance of a flapping foil for a self-propelled
  fishlike body}.  \jt{Sci. Rep.}  \bvol{11},  \pg{22297}.

\bibitem[Phoemsapthawee {\em et~al.\/}(2020)Phoemsapthawee, Thaweewat \&
  Juntasaro]{photh20}
{\sc \au{Phoemsapthawee, S.}, \au{Thaweewat, N.} \& \au{Juntasaro, V.}}
  \yr{2020}  \at{Influence of resonance on the performance of semi-active
  flapping propulsor}.  \jt{Ship Technol. Res.}  \bvol{67},  \pg{51--60}.

\bibitem[Prakash {\em et~al.\/}(2024)Prakash, Nair, Arunav, Akhil, Tawk \&
  Shankar]{prana24}
{\sc \au{Prakash, A.}, \au{Nair, A.~R.}, \au{Arunav, H.}, \au{Akhil, V.~M.},
  \au{Tawk, C.} \& \au{Shankar, K.~V.}} \yr{2024}  \at{Bioinspiration and
  biomimetics in marine robotics: a review on current applications and future
  trends}.  \jt{Bioinsp. Biomim.}  \bvol{19}~(3),  \pg{031002}.

\bibitem[Quinn {\em et~al.\/}(2015)Quinn, Lauder \& Smits]{quila15}
{\sc \au{Quinn, D.~B.}, \au{Lauder, G.~V.} \& \au{Smits, A.~J.}} \yr{2015}
  \at{Maximizing the efficiency of a flexible propulsor using experimental
  optimization}.  \jt{J. Fluid Mech.}  \bvol{767},  \pg{430--448}.

\bibitem[S\'anchez-Rodr\'iguez {\em et~al.\/}(2020)S\'anchez-Rodr\'iguez,
  Raufaste \& Argentina]{sanra20}
{\sc \au{S\'anchez-Rodr\'iguez, J.}, \au{Raufaste, C.} \& \au{Argentina, M.}}
  \yr{2020}  \at{A minimal model of self propelled locomotion}.  \jt{J. Fluids
  Structures}  \bvol{97},  \pg{103071}.

\bibitem[Smits(2019)]{smits19}
{\sc \au{Smits, A.~J.}} \yr{2019}  \at{Undulatory and oscillatory swimming}.
  \jt{J. Fluid Mech.}  \bvol{874},  \pg{P1, 1--70}.

\bibitem[Spagnolie {\em et~al.\/}(2010)Spagnolie, Moret, Shelley \&
  Zhang]{spamo10}
{\sc \au{Spagnolie, S.~E.}, \au{Moret, L.}, \au{Shelley, M.~J.} \& \au{Zhang,
  J.}} \yr{2010}  \at{Surprising behaviors in flapping locomotion with passive
  pitching}.  \jt{Phys. Fluids}  \bvol{22}~(4),  \pg{041903}.

\bibitem[Taylor {\em et~al.\/}(2003)Taylor, Nudds \& Thomas]{taynu03}
{\sc \au{Taylor, G.~K.}, \au{Nudds, R.~L.} \& \au{Thomas, A. L.~R.}} \yr{2003}
  \at{Flying and swimming animals cruise at a \text{Strouhal} number tuned for
  high power efficiency}.  \jt{Nature}  \bvol{425},  \pg{707--711}.

\bibitem[Thaweewat {\em et~al.\/}(2018)Thaweewat, Phoemsapthawee \&
  Juntasaro]{thaph18}
{\sc \au{Thaweewat, N.}, \au{Phoemsapthawee, S.} \& \au{Juntasaro, V.}}
  \yr{2018}  \at{Semi-active flapping foil for marine propulsion}.  \jt{Ocean
  Eng.}  \bvol{147},  \pg{556--564}.

\bibitem[Theodorsen(1935)]{theod35}
{\sc \au{Theodorsen, T.}} \yr{1935}  \bt{General theory of aerodynamic
  instability and the mechanism of flutter}. {\em Tech. Rep.\/} TR 496.
  \org{NACA}.

\bibitem[Triantafyllou {\em et~al.\/}(1993)Triantafyllou, Triantafyllou \&
  Grosenbaugh]{tritr93}
{\sc \au{Triantafyllou, G.~S.}, \au{Triantafyllou, M.~S.} \& \au{Grosenbaugh,
  M.~A.}} \yr{1993}  \at{Optimal thrust development in oscillating foils with
  application to fish propulsion}.  \jt{J. Fluid Structures}  \bvol{7},
  \pg{205--224}.

\bibitem[Webb(1984)]{webb84}
{\sc \au{Webb, P.~W.}} \yr{1984}  \at{Form and function in fish swimming}.
  \jt{Sci. Am.}  \bvol{251}~(1),  \pg{72--83}.

\bibitem[Willis {\em et~al.\/}(2007)Willis, Israeli, Persson, Drela, Peraire,
  Swartz \& Breuer]{wilis07}
{\sc \au{Willis, D.~J.}, \au{Israeli, E.~R.}, \au{Persson, P.~O.}, \au{Drela,
  M.}, \au{Peraire, J.}, \au{Swartz, S.~M.} \& \au{Breuer, K.~S.}} \yr{2007}
  \at{A computational framework for fluid structure interaction in biologically
  inspired flapping flight}.  \jt{AIAA Paper 2007-3803} .

\bibitem[Yao {\em et~al.\/}(2025)Yao, Xiao, Zeng, Ji, Jiang \& Shan]{yaoxi25}
{\sc \au{Yao, Q.}, \au{Xiao, S.}, \au{Zeng, Z.}, \au{Ji, Z.}, \au{Jiang, S.} \&
  \au{Shan, F.}} \yr{2025}  \at{A review of particle image velocimetry in
  fish-inspired biomimetic propulsion}.  \jt{Ocean Eng.}  \bvol{341},
  \pg{122453}.

\bibitem[Zhang {\em et~al.\/}(2010)Zhang, Liu \& Lu]{zhali10}
{\sc \au{Zhang, J.}, \au{Liu, N.~S.} \& \au{Lu, X.~Y.}} \yr{2010}
  \at{Locomotion of a passively flapping flat plate}.  \jt{J. Fluid Mech.}
  \bvol{659},  \pg{43--68}.

\bibitem[Zhong {\em et~al.\/}(2021)Zhong, Zhu, Fish, Kerr, Downs, Bart-Smith \&
  Quinn]{zhozh21}
{\sc \au{Zhong, Q.}, \au{Zhu, J.}, \au{Fish, F.~E.}, \au{Kerr, S.~J.},
  \au{Downs, A.~M.}, \au{Bart-Smith, H.} \& \au{Quinn, D.~B.}} \yr{2021}
  \at{Tunable stiffness enables fast and efficient swimming in fish-like
  robots}.  \jt{Sci. Rob.}  \bvol{6}~(57),  \pg{eabe4088}.

\bibitem[Zhu {\em et~al.\/}(2019)Zhu, White, Wainwright, Di~Santo, Lauder \&
  Bart-Smith]{zhuwh19}
{\sc \au{Zhu, J.}, \au{White, C.}, \au{Wainwright, D.~K.}, \au{Di~Santo, V.},
  \au{Lauder, G.~V.} \& \au{Bart-Smith, H.}} \yr{2019}  \at{Tuna robotics: {A}
  high-frequency experimental platform exploring the performance space of
  swimming fishes}.  \jt{Sci. Rob.}  \bvol{4}~(34),  \pg{eaax4615}.

\end{thebibliography}

\end{document}